\documentclass[twocolumn,10pt]{IEEEtran}

\usepackage{calc}

\usepackage{xcolor,multirow,gensymb}
  \usepackage{cite}
\usepackage{graphicx,subfigure,epsfig}
\usepackage{psfrag}
\usepackage{amsmath,amssymb}
\usepackage{color}
\usepackage{array}
\usepackage{nicefrac}
\usepackage{amsmath,amssymb}
\usepackage{xifthen}
\usepackage{mathtools}
\usepackage{enumerate}
\usepackage{microtype}
\usepackage{xspace}
\usepackage{bm}
\usepackage[T1]{fontenc}
\usepackage{fancyhdr}
\usepackage{lastpage}
\usepackage{rotating}
\usepackage{tabulary}

\newcommand{\vect}[1]{{\lowercase{\mbs{#1}}}}

\newcommand{\mbs}[1]{\bm{#1}}
\newcommand{\mat}[1]{{\uppercase{\mbs{#1}}}}
\newcommand{\Id}{\mat{\mathrm{I}}}

\newcommand{\T}{{\scriptscriptstyle\mathsf{T}}}
\renewcommand{\H}{{\scriptscriptstyle\mathsf{H}}}

\renewcommand{\Re}[1][]{\ifthenelse{\isempty{#1}}{\operatorname{Re}}{\operatorname{Re}\left(#1\right)}}
\renewcommand{\Im}[1][]{\ifthenelse{\isempty{#1}}{\operatorname{Im}}{\operatorname{Im}\left(#1\right)}}
\usepackage{float}

\newfloat{longequation}{b}{ext}

\newcommand{\bv}{\vect{b}}

\newcommand{\ev}{\vect{e}}

\newcommand{\etv}{\vect{\eta}}

\newcommand{\deltav}{\hbox{\boldmath$\delta$}}

\def\bA{{\mathbf{A}}}
\def\bB{{\mathbf{B}}}

\def\bD{{\mathbf{D}}}

\def\bF{{\mathbf{F}}}
\def\bG{{\mathbf{G}}}
\def\bH{{\mathbf{H}}}

\def\bJ{{\mathbf{J}}}
\def\bK{{\mathbf{K}}}
\def\bL{{\mathbf{L}}}

\def\bQ{{\mathbf{Q}}}
\def\bR{{\mathbf{R}}}
\def\bS{{\mathbf{S}}}
\def\bT{{\mathbf{T}}}

\def\bW{{\mathbf{W}}}
\def\bX{{\mathbf{X}}}

\def\bZ{{\mathbf{Z}}}
\def\bSigma{{\mathbf{\Sigma}}}

\newcommand{\cC}{{\cal C}}

\newcommand{\cN}{{\cal N}}

\def\rp{{\mathrm{p}}}

\def\rt{{\mathrm{t}}}

\def\bb{{\mathbf{b}}}

\def\bee{{\mathbf{e}}}
\def\bff{{\mathbf{f}}}
\def\bg{{\mathbf{g}}}
\def\bh{{\mathbf{h}}}

\def\bs{{\mathbf{s}}}

\def\bv{{\mathbf{v}}}
\def\bw{{\mathbf{w}}}
\def\bx{{\mathbf{x}}}
\def\by{{\mathbf{y}}}

\def\b0{{\mathbf{0}}}

\def\bTheta{{\boldsymbol{\Theta}}}

\newcommand{\Tm}{{\bf T}}

\def\bbC{{\mathbb{C}}}

\def\bpsi{{\boldsymbol{\psi}}}

\newcommand{\CC}{\mathbb{C}}

\newcommand{\EE}{\mathbb{E}}

\newcommand{\CN}[1][]{\ifthenelse{\isempty{#1}}{\mathcal{N}_{\mathbb{C}}}{\mathcal{N}_{\mathbb{C}}\left(#1\right)}}

\renewcommand{\P}[1][]{\ifthenelse{\isempty{#1}}{\mathbb{P}}{\mathbb{P}\left(#1\right)}}
\newcommand{\E}[1][]{\ifthenelse{\isempty{#1}}{\mathbb{E}}{\mathbb{E}\left(#1\right)}}
\renewcommand{\det}[1][]{\ifthenelse{\isempty{#1}}{\text{det}}{\text{det}\left(#1\right)}}
\newcommand{\trace}[1][]{\ifthenelse{\isempty{#1}}{\text{tr}}{\text{tr}\left(#1\right)}}
\newcommand{\rank}[1][]{\ifthenelse{\isempty{#1}}{\text{rank}}{\text{rank}\left(#1\right)}}
\newcommand{\diag}[1][]{\ifthenelse{\isempty{#1}}{\text{diag}}{\text{diag}\left(#1\right)}}
\def\nn{\nonumber}

\IEEEoverridecommandlockouts

\newtheorem{proposition}{Proposition}
\newtheorem{remark}{Remark}
\newtheorem{lemma}{Lemma}

\newcommand{\hatvg}{\hat{\bg}} 
 
\def\bPhi{{\boldsymbol{\Phi}}}

\newcommand{\tr}{\mathop{\mathrm{tr}}\nolimits}
\newtheorem{Theorem}{Theorem}

\newcommand{\al}{\alpha}

\newcounter{enumi_saved}
\usepackage{answers}
\Newassociation{solution}{Solution}{solutionfile}

\AtBeginDocument{\Opensolutionfile{solutionfile}[\jobname]}
\AtEndDocument{\Closesolutionfile{solutionfile}\clearpage
}

\DeclareSymbolFont{matha}{OML}{txmi}{m}{it}
\DeclareMathSymbol{\varv}{\mathord}{matha}{118}
\usepackage{amsmath}

\usepackage{eurosym}
\newcommand{\overbar}[1]{\mkern 1.5mu\overline{\mkern-1.5mu#1\mkern-1.5mu}\mkern 1.5mu}
\renewcommand{\baselinestretch}{0.93}

\begin{document}
\title{Rate-Splitting to Mitigate Residual Transceiver Hardware Impairments in Massive MIMO Systems}
\author{Anastasios Papazafeiropoulos, Bruno Clerckx, and Tharmalingam Ratnarajah,   \vspace{2mm} \\
\thanks{A. Papazafeiropoulos and T. Ratnarajah are  with the  Institute for Digital Communications (IDCOM), University of Edinburgh, Edinburgh, EH9 3JL, U.K., (email: {a.papazafeiropoulos, t.ratnarajah}@ed.ac.uk). B. Clerckx is with the Communications and Signal Processing group in the Department of Electrical and Electronic
Engineering, Imperial College London, SW7 2AZ U.K., (email: b.clerckx@imperial.ac.uk.)}
\thanks{This work was supported by the U.K. Engineering and Physical Sciences Research Council (EPSRC) under grants EP/N014073/1 and EP/N015312/1.}}
\maketitle
\begin{abstract}
Rate-Splitting (RS) has recently been shown to provide significant performance benefits in various multi-user transmission scenarios. In parallel,  the huge degrees-of-freedom provided by the appealing massive Multiple-Input Multiple-Output (MIMO)  necessitate the employment of inexpensive hardware,  being more prone to hardware imperfections, in order to be a cost-efficient technology. Hence, in this work, we focus on a realistic massive Multiple-Input Single-Output (MISO) Broadcast Channel (BC) hampered by the inevitable hardware impairments. We consider a general experimentally validated model of hardware impairments, accounting for the presence of \textit{multiplicative distortion} due to phase noise, \textit{additive distortion noise} and \textit{thermal noise amplification}.  Under both scenarios with perfect and imperfect channel state information at the transmitter (CSIT), we analyze the potential robustness of RS to each separate hardware imperfection. We analytically assess the sum-rate degradation due to hardware imperfections. Interestingly, in the case of imperfect CSIT, we demonstrate that RS is a robust strategy for multiuser MIMO in the presence of phase and amplified thermal noise, since its sum-rate does not saturate at high  signal-to-noise ratio (SNR), contrary to conventional techniques. On the other hand, the additive impairments always lead to a sum-rate  saturation at high SNR, even after the application of RS. However,  RS still enhances the performance. Furthermore, as the number of users increases, the gains provided by RS decrease not only in ideal conditions, but in practical conditions with RTHIs as well. Notably, although a deterministic equivalent analysis is employed, the analytical and simulation results coincide even for finite system dimensions. As a consequence, the applicability of these results also holds for current ``small-scale'' multi-antenna systems.   \end{abstract}

\section{Introduction}
In recent years, Multiple-Input Multiple-Output (MIMO) processing has played a central role towards the increase of both spectral and energy efficiencies~\cite{Holma2011}. Next generation systems, known as 5G (fifth generation), follow this direction, in order to cover the emerging needs from the societal development till 2020~\cite{METIS,Osseiran2014}. In fact, the advent of the Internet of Things (IoT) and the accompanied unprecedented diversity of requirements and wireless connectivity necessitate the inventiveness of enabling technologies such as massive MIMO~\cite{Marzetta2010,Rusek2013}.

As a matter of fact, massive MIMO, known also as large MIMO, is one of the backbone technologies supporting 5G systems, promising tremendous network throughput and energy efficiency~\cite{Marzetta2010,Rusek2013,Ngo2013,Larsson2014,Hoydis2013,Papazafeiropoulos2015a,Dai2016}. According to its key concept, each Base Station (BS) employs hundreds or even thousands of antennas under simple coherent processing serving an order of magnitude fewer User Elements (UEs)~\cite{Rusek2013}. Based on the law of large numbers, fast fading, intra-cell interference, and additive Gaussian noise are averaged out in the large number of antennas limit.

In practice, besides these channel impairments, inevitable imperfections, emerging from the transceiver hardware, distort both the transmit and receive signals~\cite{Schenk2008,Studer2010}. These include the In-phase/Quadrature-phase (I/Q)-imbalance~\cite{Qi2010}, the high power amplifier non-linearities~\cite{Qi2012},  and the oscillator Phase Noise (PN)~\cite{Mehrpouyan2012,Pitarokoilis2015,Krishnan2015,Papazafeiropoulos2016 }.  In real world applications, the hardware impairments can be partially mitigated by means of calibration schemes and compensation algorithms  at the transmitter and the receiver, respectively~\cite{Studer2010}. However, a certain amount of distortions remains, which results in Residual Transceiver Hardware Impairments (RTHIs). These RTHIs can be categorized in additive and multiplicative distortions. The additive impairments describe the aggregate effect of many impairments and are modeled as additive distortion noises at both the BS and UEs~\cite{Studer2010,Bjoernson2013}. In particular, the adoption of this model is grounded due to its analytical tractability and experimental validation~\cite{Studer2010}. On the other hand, phase drifts from Local Oscillators (LOs) present a multiplicative nature with regards to the channel vector. If the variation of these impairments can be assumed sufficiently static, they can be incorporated into the channel by means of an appropriate scaling of its covariance matrix. Otherwise, if they accumulate within the channel coherence time, e.g., the PN, they cannot be incorporated into the channel vector~\cite{Krishnan2015,Papazafeiropoulos2016}. Unfortunately, RTHIs are a fundamental bottleneck toward the realistic evaluation as well as the promising spectral
and energy efficiencies  benefits of  5G systems because they cannot be estimated with infinite precision. Actually, not only they constitute a degradation source of the system performance, but they also result in an inaccurate Channel State Information at the Transmitter (CSIT) which degrades even further the spectral efficiency. 

The study of the impact of hardware impairments has originated from conventional wireless systems and has continued to 5G networks by means of massive MIMO systems~\cite{Schenk2008,Goransson2008,Papazafeiropoulos2016,Qi2010,Qi2012,Mehrpouyan2012,Studer2010,Bjoernson2013,Zhang2014,Bjornson2014,Krishnan2015,Bjornson2015,Pitarokoilis2015,A.K.Papazafeiropoulos2015,PapazafeiropoulosMay2016,PapazafeiropoulosJuly2016} and heterogeneous networks design\cite{PapazafComLetter2016}. Especially, the deployment of massive MIMO systems is attractive if the antenna elements consist of inexpensive hardware components. Unfortunately, the majority of massive MIMO literature has assumed perfect transceiver hardware, although hardware imperfections exist.  Reasonably, it is conjectured that following the same path will increase the gap between theory and practice. Hence, misleading conclusions could be made during the design and evaluation of the forthcoming 5G systems.

At the same time, obtaining accurate  CSIT is a challenging task, especially, as the number of BS antennas increases. In such case, Time Division Duplex (TDD) architectures have proved to be a more feasible solution against Frequency Division Duplex (FDD) designs that are accompanied with channel estimations and feedback challenges~\cite{Marzetta2010,Rusek2013,Ngo2013,Larsson2014,Hoydis2013}.

Herein, we tackle the challenge of mitigating the RTHIs in massive  Multiple-Input Single-Output (MISO). Specifically, we leverage the Rate-Splitting (RS) approach, where we can split one's UE message into a common part and a private part~\cite{Clerckx2016,Zappone2015,Hao2017,Joudeh2016b}. The common message, drawn from a public codebook, can be decoded by all UEs with zero error probability, while  the private messages are transmitted by means of linear beamforming such as Zero-Forcing (ZF). Each private message occupies a fraction of the total power, while the common message, superimposed on top of the private messages uses the residual power. In the practical case of imperfect CSIT, if the error variance is fixed, linear precoding techniques lead to a rate ceiling at high Signal-to-Noise Ratio (SNR) due to multi-user interference.  Interestingly,  the RS outperforms conventional broadcasting  at high SNR, since it  does not experience any ceiling effect~\cite{Hao2015,Dai2016}. Henceforth, we denote by NoRS all the  conventional techniques to contrast with the RS techniques. A further gain of RS over NoRS can be achieved by optimizing the precoders~\cite{Joudeh2016,Joudeh2016a}. Given that the additive RTHIs are power-dependent, while the PN is independent, we aim to exploit the RS design and illustrate its potential robustness  in massive MIMO systems by means of a Deterministic Equivalent (DE) analysis~\cite{Hoydis2013,Papazafeiropoulos2015a,Papazafeiropoulos2016,Krishnan2015}. 

This paper investigates the potential robustness of the RS approach in realistic  massive MISO systems with RTHIs in both cases of perfect and imperfect CSIT  implemented under TDD architectures. Actually, the source of imperfect CSIT is the pilot contamination, while the RTHIs contribute further to this imperfection. In other words, when perfect CSIT is accounted, it reflects the channel obtained by the ideal scenario of no channel estimation error and no RTHIs on the uplink. Moreover, we perform our analysis for two different settings. In the first setting, all BS antennas are connected to a Common LO (CLO), while the other design refers to BS antennas with Separate LOs (SLOs). 

\subsection{Motivation-Central Idea}
The paper is motivated by the following two observations: 1) RS  tackles efficiently multi-user interference, and in particular the one arising due to the imperfect CSIT in MISO BC, 2) CSI is effectively altered due to the presence of RTHIs, e.g., PN induces a fast variation of the channel between the channel estimation time and the actual data transmission. These observations suggest that RS may be a suitable alternative to conventional strategies in order to mitigate RTHIs.
%
The main contributions are summarized as follows.
\begin{itemize}
 \item We shed light on the impact of RTHIs on the performance of the downlink of a TDD-based Massive MISO system with RS and NoRS. Specifically, we take into account   multiplicative and additive impairments, as well as  amplified thermal noise in the general scenario where imperfect CSIT is available. For the sake of comparison, we also present the results corresponding to perfect CSIT. 
 \item Contrary to existing works~\cite{Bjornson2015,Krishnan2015} which have studied the effect of RTHIs on
the performance degradation of the uplink, we focus on the MISO downlink with also a large number of BS antennas,  and examine the impact of various impairments. Herein, it should be stressed that uplink RTHIs impact the channel estimation stage, which also has an impact on the DL performance.  
 \item We derive the deterministic signal-to-interference-plus-noise ratios (SINRs) of NoRS and RS with RTHIs and use them to investigate the performance benefits of RS over NoRS in the presence of RTHIs. Specifically, first, we obtain the estimated channel by means of MMSE estimation. Next, we design the precoder of the  private and common messages, and we perform suitable power allocation. At the end, we provide the DEs of the SINRs  of the  private and common messages\footnote{This work employs Regularized ZF (RZF) precoding for the transmission of the private messages, although the basic implementation of the RS method involves just ZF precoding except~\cite{Dai2016}. However, the robustness of RZF and its lack in the literature while investigating RTHIs led us to enroll it.}. These deterministic expressions allow to avoid any Monte Carlo simulations with very high precision. 
 \item We elaborate on the impact of each separate impairment on the sum-rate performance with  RS and NoRS. Remarkably, RS outperforms the NoRS strategies at high SNR in the cases where only phase and amplified thermal noises are assumed. 
 Actually, RS is able to mitigate the saturation due to  the unavoidable PN in spite of the knowledge of perfect or imperfect CSIT. ln the case where only additive hardware impairments are present, the saturation is inevitable even with the implementation of RS. However, RS still provides some SNR gains over NoRS.
\item Increasing the number of UEs results in a reduction of the performance gain of RS over NoRS because the common message has to be decoded by more UEs. We quantify this degradation in the presence of RTHIs.
 \end{itemize}

The   remainder of this paper is structured as follows.  Section~\ref{SystemModel} presents the system  and signal models for a  BC system with multiple transmit antennas and ideal hardware under the NoRS and RS approaches. In Section~\ref{HI}, we present the various impairments (both additive and multiplicative) under consideration. Next, in Section~\ref{training}, we provide the estimated channel obtained during the uplink training phase with RTHIs, while Section~\ref{Downlink} shows the downlink transmission under hardware impairments.  Section~\ref{Deterministic} exposes the DE analysis, which enables the design of the  precoder of the common message, and mainly, the derivation of the achievable rates in the presence of RTHIs.  For comparison, Section~\ref{DeterministicPerfect} presents briefly the corresponding results by assuming perfect CSIT. The numerical results are placed in Section~\ref{NumericalResults}, while Section~\ref{Conclusions} summarizes the paper.

\textit{Notation:} Vectors and matrices are denoted by boldface lower and upper case symbols. $(\cdot)^\T$, $(\cdot)^*$,  $(\cdot)^\H$, and $\tr\!\left( {\cdot} \right)$ represent the transpose, conjugate, Hermitian  transpose, and trace operators, respectively. The expectation  operator is denoted by $\EE\left[\cdot\right]$. The $\mathrm{diag}\{\cdot\}$ operator generates a diagonal matrix from a given vector, and the symbol $\triangleq$ declares definition. The notations $\mathcal{C}^{M \times 1}$ and $\mathcal{C}^{M\times N}$ refer to complex $M$-dimensional vectors and  $M\times N$ matrices, respectively. Finally, $\bb \sim \cC\cN{(\b0,\mathbf{\Sigma})}$ and $\bb \sim \cN{(\b0,\mathbf{\Sigma})}$ denote a circularly symmetric complex Gaussian variable with zero-mean and covariance matrix $\mathbf{\Sigma}$ and the corresponding real Gaussian variable, respectively.



\section{System Model}\label{SystemModel} 
In this paper, we consider a MISO BC channel comprising a $M$-antenna BS and $K$ single-antenna UEs forming a set with cardinality $\mathcal{K}$. Especially, the transmitter antennas are co-located at a macro BS  that communicates simultaneously with $K$ receivers.   

The frequency-flat channel between the BS and UE $k$, modeled as Rayleigh block fading\footnote{It is  static across a coherence block of $T$ channel uses, while the channel realizations between blocks are independent. The size of the block depends on the the product of the coherence time $T_{c}$ and the coherence bandwidth $B_{c}$. For example, if $T_{c}=2$ ms and $B_{c}=100$ KHz, then $T=200$ channel uses.  }, is denoted by $\bh_{k}\triangleq \left[h_{k}^{1}, \ldots, h_{k}^{M}  \right]\in \bbC^{M\times 1}$. We express each channel realization as
\begin{align}
\bh_{k} = \bR^{1/2}_{k}\bw_{k},
\end{align}
where $\bR_{k}\! =\!\mathbb{E}\!\left[ \bh_{k} \bh_{k}^{\H}\right]\!\in\! \bbC^{M \times M }$ is a deterministic Hermitian-symmetric positive-definite matrix representing versatile effects such as the assignment of antenna correlation to each channel vector or different path loss to each antenna.  Regarding $\bw_{k} \in \bbC^{M \times 1}$, it is an uncorrelated fast-fading Gaussian channel vector drawn as  $\bw_{k} \sim \cC\cN(\b0,\Id_{M})$. In other words, we can write \begin{align}
 \bh_{k} \sim \cC\cN \left( \b0,\bR_{k} \right).
\end{align}

\subsection{Conventional Approach (NoRS)}
Let us first present a conventional MISO BC with a linear precoder. The  signal received by UE $k$  can be written as
\begin{align}
y_{k}=\bh_{k}^{\H} \bx+ z_{k}, \label{BasicSystemModelDownlinkwithoutRTHIs}
\end{align}
where $z_{k} \sim \cC\cN(0,1)$ is the Additive White Gaussian Noise (AWGN) at UE $k$ and
\begin{align}
 \bx&=\sum_{k=1}^{K}\sqrt{\lambda\tilde{\rho}_{k}} \bff_{k} s_{k},\nn\\
 &=\sqrt{\lambda \tilde{\rho}_{k}} \bF \bs\label{transmit}
\end{align}
is the transmit signal with   $\tilde{\rho}_{k}=\rho_{k}$ being the downlink transmit power (equal power allocation for all users)\footnote{The uniform power allocation is commonly used in the Massive MIMO literature~\cite{Marzetta2010,Ngo2013,Dai2016}. This assumption is used primarily as a simplification, but it is also motivated by practical deployments of MU-MIMO, where uniform power allocation is actually used in 4G deployment, as explained in e.g.~\cite{Lim2013}. Further enhancement could be obtained by jointly optimizing the power allocation as well as the precoders of the common and private messages. This approach would lead to the optimal precoders and has been used for rate splitting with imperfect CSIT (and no hardware impairments) in~\cite{Joudeh2016}. However it is not really practical to resort to this type of optimization for large-scale systems such as Massive MIMO, where the use of deterministic equivalent is commonly used in order to get some further insight into the system behaviour in terms of different aspects such as the impact of hardware impairments.},  $s_{k}$ being the private message for UE $k$, and $\bff_{k}$ being the linear precoder corresponding to UE $k$, which is designed as RZF. In addition, $\bF \in \bbC^{M \times K} $ is the precoder multiplying the data symbol vector  $\bs \in \bbC ^{K \times 1}\sim \cC\cN \left( \b0,\Id_{K} \right)$. Note that $\lambda$ is the normalization parameter regarding the precoder given by
\begin{align}
\lambda=\frac{K}{\EE \left[ \tr\bF^\H\bF \right]}. \label{eq:lamda} 
\end{align}

\subsection{RS Approach}
The description above concerns a conventional linearly precoded multi-user broadcasting scheme. Given that our focal point is to investigate the performance  of the promising RS method under the  unavoidable RTHIs, we provide shortly its presentation. 

In~\cite{Hao2015,Dai2016,Clerckx2016,Zappone2015,Hao2017,Joudeh2016b,Joudeh2016,Joudeh2016a}, RS was shown to be very promising in multi-user transmission with imperfect CSIT. It indeed achieves unsaturated sum-rate with increasing SNR despite the presence of imperfect CSIT. Contrary to the NoRS strategy that treats as noise any multi-user interference originating from the imperfect CSIT, the RS strategy is able to bridge treating interference as noise and perform interference decoding through the presence of a common message. This ability of the decoding part of the interference is the key  boosting the sum-rate performance. Motivated by this observation, RS is also expected to provide benefits in the presence of some RTHIs, since RTHIs have the effect of altering the CSI between the estimation stage and the transmission stage. 

The basic principle of this method imposes the message intended for UE $k$ to be split into two parts, namely, a common and a private part. The  common part, drawn from a public codebook,  should be decoded by all UEs with
zero error probability. On the other hand, the private part is to be decoded only by UE $k$. Regarding the messages intended for the other UEs, these consist of a private part only. 
Mathematically speaking,~\eqref{transmit} becomes
\begin{align}
 \bx=\underbrace{\sqrt{ \rho_{\mathrm{c}}}\bff_{c} s_{c}}_{\mathrm{common ~part}}+\underbrace{\sum_{k=1}^{K}\sqrt{\lambda \rho_{{k}}}\bff_{k} s_{k}}_{private~part},\label{RStransmit}
\end{align}
where $s_{c}$ is the common message and $s_{k}$ is the private message of UE $k$, while $\bff_{c}$ denotes the  precoding vector of the common
message with unit norm and $\bff_{k}$ is the linear precoder corresponding to UE $k$. Moreover, $\rho_{\mathrm{c}}$ is the power allocated to the common signal. In other words, the private messages $s_{k} \forall k$ are superimposed over the common message $s_{c}$ and sent with linear precoding. We assume that the total power budget of RS and NoRS is the same, i.e., $\tilde{\rho}_{k}=\rho_{c}+\rho_{k}$. It is worthwhile to present the decoding procedure. First, the common message is decoded by each UE, while all private messages are treated as noise. Next, each UE subtracts the contribution of the common message from the received signal and is able to decode its own private message. Herein, we focus on the application of the RZF precoder for the private messages, as mentioned before. Also, hereafter, we are going to use the notion of the time slot in our expressions, since the hardware impairments, presented below, are time dependent.

\section{Hardware Impairments}\label{HI} 
In practice, both  the transmitter and the receiver are affected by various impairments detailed below. Specifically, we present the models describing  1) the multiplicative PN at both the transmitter and the receiver, 2) the additive power-dependent distortion noises at the transmitter and receiver, and 3) the amplified thermal noise at the receiver side. Hereafter, we assume that the hardware impairments parameters are assumed to be known by their manufacturer by means of certain specifications.
\subsection{Phase Noise}\label{PN1} 
The PN, being the distortion in the phase due to the random phase drift  in the signal coming from the LOs of the BS and UE $k$, is induced during the up-conversion of the baseband signal to passband and vice-versa\footnote{The conversion takes place by multiplying the signal with the LO's output.}.

According to~\cite{Demir2000,Pitarokoilis2015}, the PN during the $n$th time slot can be described by a discrete-time  independent Wiener process, i.e., the PNs at the  LOs of the $m$th antenna of the BS and $k$th UE are modeled as 
\begin{align}
 \phi_{m,n}&=\phi_{m,n-1}+\delta^{\phi_{m}}_{n}\label{phaseNoiseBS}\\
 \varphi_{k,n}&=\varphi_{k,n-1}+\delta^{\varphi_{k}}_{n},\label{phaseNoiseuser}
\end{align}
where $\delta^{\phi_{m}}_{n}\sim \cN(0,\sigma_{\phi_{m}}^{2}) $ and $\delta^{\varphi_{k}}_{n}\sim \cN(0,\sigma_{\varphi_{k}}^{2})$. Note that $\sigma_{i}^{2}=4\pi^{2}f_{\mathrm{c}} c_{i}T_{\mathrm{s}}$, $i=\phi_{m}, \varphi_{k}$ describes the PN increment variance with $T_{\mathrm{s}}$, $c_{{i}}$, and $f_{\mathrm{c}}$ being the  symbol interval, a constant dependent on the oscillator, and the carrier frequency, respectively.

We assume non-synchronous operation at the BS, if the BS antennas have independent PN processes $\phi_{m,n},~ m=1,\ldots,M$ with $\phi_{m,n}$ being the PN process at the $m$th antenna. Note that the  PN processes are considered as mutually independent, since each antenna has its own oscillator, i.e., an  SLO at each antenna, while when  the PN processes $\phi_{m,n}$ are identical for all $m = 1, \ldots, M$ we have the synchronous operation.  In case that we have just one CLO connected to all BS antennas, there is only one PN process $\phi_{n}$. In our analysis, we focus  on both SLOs and CLO scenarios, but in all cases, we assume  i.i.d. PN statistics across different antennas and UEs, i.e., $\sigma_{\phi_{m}}^{2}=\sigma_{\phi}^{2}$ and $ \sigma_{\varphi_{k}}^{2}=\sigma_{\varphi}^{2}~,\forall~m,~k$.

The PN is expressed as a multiplicative factor to the channel vector as 
\begin{align}
\tilde{\bg}_{k,n}=\bTheta_{k,n}\bh_{k}\label{PhaseNoise} ,
\end{align}
where $\bTheta_{k,n}\!\triangleq\!\mathrm{diag}\!\left\{ e^{j \theta_{k,n}^{(1)}}, \ldots, e^{j \theta_{k,n}^{(M)}} \!\right\}=e^{j \varphi_{k,n}}\bPhi_{n}\in \mathbb{C}^{M\times M}$ is the total PN with $\bPhi_{n}\!\triangleq\!\mathrm{diag}\!\left\{\! e^{j \phi_{1,n}}, \ldots, e^{j \phi_{M,n}} \!\right\}$ being the PN sample matrix  at time $n$ because of the imperfections in the LOs of the BS. Similarly, $e^{j \varphi_{k,n}}$ is the PN induced by UE $k$. Basically, $\tilde{\bg}_{k,n}$ represents the effective channel vector at time $n$. Interestingly, the effective channel, given by \eqref{PhaseNoise}, depends on the time slot of symbol $n$ due to the time-dependence coming from the PN. 
\subsection{Additive Distortion Noise}\label{ADN} 
In real systems, both the transmitter and the receiver are affected by  inevitable residual additive impairments that emerge after imperfect compensation of the quantization noise in the Analog-to-Digital Converters (ADCs) at the receiver, the I/Q imbalance, etc.~\cite{Schenk2008}. As a result, the received signal is distorted during the reception processing, while at the transmitter side, a mismatch appears between the signal that  is intended to be transmitted and the generated signal. 

Let $T_\mathrm{i}$ and $R_\mathrm{j}$ be the numbers of transmit and receive antennas of nodes $i,~j$ depending on their role, i.e., if node $i$ is the UE or the BS in transmit mode, we have  $T_\mathrm{UE}=1$ or $T_\mathrm{BS}=M$. Correspondingly, if node $j$ is the BS or the UE in receive mode, we have  $R_\mathrm{BS}=M$ or  $R_\mathrm{UE}=1$, respectively.  $\bQ_{i}$ is the transmit covariance matrix of the corresponding node with diagonal elements $q_{\mathrm{i}_1},\ldots,q_{T_\mathrm{i}}$, e.g., if the transmitter node is the UE, $\bQ_{i}$ degenerates to a scalar $Q_{\mathrm{UE}}$. 

Generally, the transmitter and  receiver distortion noises  are modeled as Gaussian distributed, where their average power is proportional to the average signal power, as shown by measurement results~\cite{Studer2010,Wenk2010}.  The circularly-symmetric complex Gaussianity can be justified by the aggregate contribution of many impairments~\cite{Schenk2008,Bjornson2014}.  Note that other impairments such as antenna coupling or even in-phase and quadrature imbalance,  attenuating the amplitude and rotating the phase of the desired constellation, cannot be modeled exactly by the residual additive impairments; however, they can be described roughly by means of their aggregate contribution. Mathematically speaking, we have
\begin{align}
 \etv_{\mathrm{t},n}^{\mathrm{i}}&\sim \cC\cN\left( \b0,\bm \Lambda^{\mathrm{i}}_{n} \right)\label{eta_t} \\
 \etv_{\mathrm{r},n}^\mathrm{j}&\sim \cC\cN \left( \b0,\bm \Upsilon^{\mathrm{j}}_{n} \right)\label{eta_r},
\end{align}
where $\bm \Lambda^{\mathrm{i}}_{n}= \kappa_{\mathrm{t}_\mathrm{i}}^{2}\mathrm{diag}\left( q_{1},\ldots,q_{T_\mathrm{i}} \right)$ and 
$\bm \Upsilon^{\mathrm{j}}_{n} =\kappa_{\mathrm{r}_{\mathrm{i}}}^{2}\sum_{k=1}^{|j |}\bh_{k,n}^{\H}\bQ_{k}^{j}\bh_{k,n} $. Note that if $j=\mathrm{UE}$, then $|j|=1$\footnote{The additive distortions with   quantization noise, being a main cause,  are time-dependent because they take new realizations for each new data signal.}. Otherwise, if  $j=\mathrm{BS}$, then $|j|=K$. 
Furthermore, $\kappa_{\mathrm{t}_\mathrm{i}}^{2}$ and $\kappa_{\mathrm{r}_\mathrm{i}}^{2}$ are proportionality parameters describing the severity of the residual additive impairments at the transmitter and the receiver of link $i$, and basically, express the  ratio between the additive distortion noise variance and the signal power. In practical applications, these parameters appear as the Error Vector Magnitudes (EVM) at each transceiver side~\cite{Holma2011}.
\subsection{Amplified Thermal  Noise}\label{TN} 
The low noise amplifier, the mixers at the receiver  as well as other components engender an amplification of the thermal noise, which appears as an increase of its variance~\cite{Bjornson2015}. In fact, the total effect $\bm \xi_{n}^{\mathrm{i}}$ can be modeled as  Gaussian distributed with zero-mean and variance $ \xi_{n}^{\mathrm{i}} \Id_{R_{\mathrm{i}}}$, where  $ \sigma^{2}\le \xi_{n}^{\mathrm{i}} $ is the corresponding parameter of the actual thermal noise\footnote{The thermal noise also takes different random  realizations over time, since it is constituted of  samples from a white noise process that has passed by some amplified ``filter''.}.
\begin{remark}
The conventional ideal model with no hardware impairments is obtained if $\sigma_{\phi_{m}}=\sigma_{\varphi_{k}}=\kappa_{\mathrm{t}_\mathrm{i},n}=\kappa_{\mathrm{r}_\mathrm{i},n}=0$, and $\bm \xi_{n} = \sigma^{2}$  $\forall m,~k,~i,~n$\footnote{Note that among the effects that are modeled by the  additive impairments  are the in-phase and quadrature imbalance, while the amplified thermal noise can model the amplifier nonlinearities.}.
\end{remark}

\section{Uplink Pilot Training Phase with RTHIs}\label{training} 
In the case of imperfect CSIT, the transmission protocol, assuming TDD, includes coherence blocks, where each one has a duration of $T$ channel uses and is split into  uplink pilot symbols and downlink data symbols. Actually, $\tau \ge K$ symbols are allocated  for pilot transmission to guarantee that the UEs are spatially separable by the corresponding BS, i.e., to avoid intra-cell interference. The remaining $T-\tau$ channel
uses are dedicated for data transmission. The CSI is  acquired during the uplink training phase, while the downlink channel is known by exploiting
the property of channel reciprocity. During this phase, a predefined pilot sequence of $\tau$ symbols is assigned to  UE $k$, i.e., $\bm \omega_{k}\triangleq \left[\omega_{k,1},\ldots,\omega_{k,\tau} \right]^{\T}\in \bbC^{\tau \times 1}$ with $\rho_{up}^{\mathrm{UE}}=\left[|\omega_{k,n}|^{2} \right],\forall k,n$, while the sequences among different UEs are mutually orthogonal.

The received uplink vector at the BS at time $n \in \left[0, \tau \right]$ $\by^{\mathrm{tr}}_{n} \in \bbC^{M \times 1}$, accounting for the RTHIs, is given by 
\begin{align}
&\!\!\!\by^{\mathrm{tr}}_{n}\!=\!\sum_{k=1}^{K}\tilde{\bg}_{k,n}\left(\omega_{k,n}+\eta_{\mathrm{t},n}^{\mathrm{UE}} \right)\!+\!\etv_{\mathrm{r},n}^{\mathrm{BS}}+\bm \xi_{n}^{\mathrm{BS}} \label{BasicSystemModel},
\end{align}
where  $\eta_{\mathrm{t},n}^{\mathrm{UE}}$,  $\etv_{\mathrm{r},n}^{\mathrm{BS}}$, and  $\bm \xi^{\mathrm{BS}}_{n}$ denote the  distortion noises at the transmitter (UE) and receiver (BS), and the amplified thermal noise  at time instance $n$.  As mentioned, $\bh_{k}$ is assumed to be constant during the coherence time $T$, while it  changes independently afterwards.  Based on~\eqref{eta_t} and~\eqref{eta_r} for $\mathrm{i}=\mathrm{UE}$ and $\mathrm{j}=\mathrm{BS}$, respectively, we have
\begin{align}
\Lambda^{\mathrm{UE}}_{n}&=\kappa_{\mathrm{t}_\mathrm{UE}}^{2}\rho_{up}^{\mathrm{UE}}\\
\bm \Upsilon^{\mathrm{BS}}_{n}&=\kappa_{\mathrm{r}_\mathrm{BS}}^{2} \rho_{up}^{\mathrm{UE}}\sum_{k=1}^{K}  \mathrm{diag}\left( |h_{k}^{1}|^{2},\ldots,|h_{k}^{M}|^{2} \right),
\end{align}
while $\bm \xi^{\mathrm{BS}}_{n} \sim \cC\cN \left( \b0,\xi_{n} \Id_{M} \right)$.

Concatenation of all the received signal vectors during the training phase results in a new vector $\bpsi \triangleq \left[{\by^{\mathrm{tr}}_{0}}^{\T}, \ldots, {\by^{\mathrm{tr}}_{\tau}}^{\T} \right]^{\T} \in \bbC^{\tau M\times 1} $. Similar to~\cite{Bjornson2015}, the  Linear Minimum Mean-Square Error (LMMSE) estimate of the channel of UE $k$ during the training phase is given by
\begin{align}
\hat{\bg}_{k,n}&=\EE\left[\tilde{\bg}_{k,n}\bm \psi^{\H} \right]\left( \EE\left[\bm \psi \bm \psi^{\H} \right]  \right)^{-1}\bm \psi\nn\\
&= \left( \bm \omega_{k}^{\H}\bm \Delta_{k,n}^{\mathrm{\tr}}\otimes \bR_{k}\right)\bm \Sigma^{-1}\bm \psi\label{EstimatedChannel} ,
\end{align}
where 
\begin{align}
 \bm \Delta_{k,n}^{\mathrm{\tr}}&\triangleq\mathrm{diag}\!\left\{ e^{-\frac{\sigma_{\varphi}^{2}+\sigma_{\phi}^{2}}{2}n}, \ldots, e^{-\frac{\sigma_{\varphi}^{2}+\sigma_{\phi}^{2}}{2}|n-\tau|} \!\right\}\\
 \bm \Sigma&\triangleq \sum_{j=1}^{K}\bX_{j}\otimes\bR_{j}+\xi^{\mathrm{BS}}\Id_{\tau M},\\
 \bX_{j}&\triangleq \tilde{\bX}_{j}+\kappa_{\mathrm{r}_\mathrm{BS}}^{2}\bD_{|\bm \omega_{j}|^{2}},\\
 \bD_{|\bm \omega_{j}|^{2}}&\triangleq \mathrm{diag}\left( |\omega_{j,1}|^{2},\ldots,|\omega_{j,\tau}|^{2} \right),\\
 \left[ \tilde{\bX}_{j}\right] _{u,v}&\triangleq \left(\omega_{j,u}\omega_{j,v}^{*}+\kappa_{\mathrm{t}_\mathrm{UE}}^{2}\right)\rho_{up}^{\mathrm{UE}}e^{-\frac{\sigma_{\varphi}^{2}+\sigma_{\phi}^{2}}{2}|u-v|}.
\end{align}
\begin{proof}
The proof, following the same steps with Theorem 1 in~\cite{Bjornson2015} by means of some algebraic manipulations, is omitted for the sake of limited space\footnote{Our expression is more general than the corresponding estimated channel in~\cite{Bjornson2015} because it includes also the transmit additive distortion, which is indirectly statistically dependent on the channels.}.
\end{proof}

Decomposing the current channel by means of the property of orthogonality of LMMSE estimation, we have that the current channel at the end of the training phase is given by 
\begin{align}
\tilde{\bg}_{k,\tau}=\hat{\bg}_{k,\tau}+\bee_{k,\tau},\label{current}                                    
\end{align}
where $\bee_{k,\tau}$ is the Gaussian distributed zero-mean estimation error vector with covariance given by\footnote{Hereafter, the subscript $\tau$ is absorbed and~\eqref{current} becomes $\tilde{\bg}_{k}=\hat{\bg}_{k}+\bee_{k}$.}
\begin{align}
\tilde{\bR}_{k}=\bR_{k}-\hat{\bR}_{k}.
\end{align}

We have $\hat{\bg}_{k}\!\sim\! \cC\cN \left( \b0,\hat{\bR}_{k} \right)$ with $\hat{\bR}_{k}=\left( \bm \omega_{k}^{\H}\bm \Delta_{k}^{\mathrm{\tr}}\otimes \bR_{k}\right)\bm \Sigma^{-1}$ $ \left( {\bm \Delta_{k}^{\mathrm{\tr}}}^{\H } \bm \omega_{k} \otimes  \bR_{k} \right)$.
\begin{remark}
In the case of ideal hardware, \eqref{EstimatedChannel} simplifies to~\cite{Hoydis2013}
\begin{align}
\hat{\bg}_{k}&=\left(  \Id_{M}+\frac{\sigma^{2}}{\rho_{\rp}}\bR_{k}^{-1}\right)^{-1}\bm \psi\label{EstimatedChannel5},
\end{align}
where $\sigma^{2}$ is the variance of the post-processed noise at  the 
base station and $\rho_{\mathrm{p}}=\tau \rho_{up}^{\mathrm{UE}}$.
\end{remark}

Comparing \eqref{EstimatedChannel} with \eqref{EstimatedChannel5}, we deduce that although the pilots are orthogonal, the phase noise induces an inherent pilot contamination, since the estimated channel  of UE $k$ is affected by means of the pilot transmissions from other UEs, as can be shown by~\eqref{EstimatedChannel}\footnote{The pilots can be ``spatially orthogonal'' or ``temporally orthogonal''. In the case of spatial orthogonality, all UEs transmit at every pilot transmission time, which effectively increases the total pilot energy by a factor $K$. When ``each UE transmits pilot signals in TDD mode with no other UE transmitting at the same time'', it is a different setting, where you get rid of the extra pilot contamination, but you also lose a factor $K$ in total pilot energy. Depending on the scenario, the reduced pilot contamination or the reduced pilot energy might dominates~\cite{Bjornson2015}.}. Furthermore, the dependence of the estimated channel on time $n$ necessitates a continuous computation of the applied precoder in the downlink at every  symbol interval, which is computationally prohibitive due to its complexity. Therefore, we assume  that the precoder is designed by means of the channel estimate once during the training phase
and then it is applied for the whole duration of the downlink transmission phase. For example, if the channel is estimated at $n_{0}=\tau$, the applied precoder is denoted by $\bff_{k  }\triangleq\bff_{k,n_{0}+1}$.
\section{Downlink Transmission under RTHIs}\label{Downlink}
Given that TDD is based on channel reciprocity, the received signal by UE $k$ during the transmission phase $n \in \left[ \tau+1, T\right] $ is given by
\begin{align}
y_{k,n}={\bh}_{k}^{\H}{\bTheta}_{k,n}^{*}\left( \bx+\etv_{\mathrm{t},n}^{\mathrm{BS}} \right)\!+\!\eta_{\mathrm{r},n}^{\mathrm{UE}}+ \xi_{k,n}^{\mathrm{UE}}, \label{BasicSystemModelDownlink}
\end{align}
where $\etv_{\mathrm{t},n}^{\mathrm{BS}}\sim \cC\cN\left( \b0,\bm \Lambda^{\mathrm{BS}} \right)$ and $\eta_{\mathrm{r},n}^{\mathrm{UE}}\sim \cC\cN \left( \b0, \Upsilon^{\mathrm{UE}} \right)$ are the residual downlink additive Gaussian distortions at the BS and the UE, which are given by~\eqref{eta_t} and~\eqref{eta_r} for $\mathrm{i}=\mathrm{BS}$ and $\mathrm{j}=\mathrm{UE}$, respectively. Specifically, we have
\begin{align}
 \bm \Lambda^{\mathrm{BS}}_{n}&=\kappa_{\mathrm{t}_\mathrm{BS}}^{2} \mathrm{diag}\left( q_{1},\ldots,q_{M} \right)\\
  \Upsilon^{\mathrm{UE}}_{n}&=\kappa_{\mathrm{r}_\mathrm{UE}}^{2}\bh_{k,n}^{\H}\bQ_{\mathrm{BS}}\bh_{k,n}.
\end{align}

Note that $\xi_{k,n}^{\mathrm{UE}}$ expresses the  amplified thermal noise at the UE $k$ at time $n$, while we assume that all UEs  present the same impairments, i.e., $\kappa_{\mathrm{r}_\mathrm{UE}}$, and $\xi_{k}^{\mathrm{UE}}$ are identical for all UEs served by the corresponding BS. Furthermore, during the downlink transmission phase described by~\eqref{BasicSystemModelDownlink}, we set $
 {\bh}_{k}^{\H}{\bTheta}_{k,\tau}^{*}= \tilde{\bg}^{\H}_{k,\tau}$. If we solve with respect to ${\bh}_{k}^{\H}$ and make the necessary substitution, we result in 
\begin{align}
 {\bh}_{k}^{\H}{\bTheta}_{k,n}^{*}= \tilde{\bg}_{k,\tau}^{\H}\widetilde{\bTheta}_{k,n}.
\end{align}
where $\widetilde{\bTheta}_{k,n}\!\triangleq\!\mathrm{diag}\!\left\{ e^{-j \left( \theta_{k,n}^{(1)}-\theta_{k,\tau}^{(1)} \right)}, \ldots, e^{-j \left( \theta_{k,n}^{(M)}-\theta_{k,\tau}^{(M)} \right)}\!\right\}$. Thus, if we set $\bg_{k,n}= \widetilde{\bTheta}_{k,n}^{*}\tilde{\bg}_{k,\tau}$,~\eqref{BasicSystemModelDownlink} becomes
\begin{align}
y_{k,n}={\bg}_{k,n}^{\H}\left( \bx+\etv_{\mathrm{t},n}^{\mathrm{BS}} \right)\!+\!\eta_{\mathrm{r},n}^{\mathrm{UE}}+ \xi_{k,n}^{\mathrm{UE}}. \label{BasicSystemModelDownlink1}
\end{align}
\begin{remark}
The term $\widetilde{\bTheta}_{k,n} $ characterizes the impact of phase noise between the training and the data transmission phases. Its trace ${T}_{\mathrm{PN}}$ is given by
\begin{align}
{T}_{\mathrm{PN}}&=\tr \widetilde\bTheta_{k,n}\nn\\
&=\sum_{l=1}^{M}e^{-j \left(  \theta_{k,n}^{(l)}-\theta_{k,\tau}^{(l)} \right)}\label{PNtrace}.
\end{align}
\end{remark}
From~\eqref{PNtrace}, we get the following lemma. 
\begin{lemma}\label{SetupLOs}
For CLO and SLOSs, we have 
\begin{align}
 \frac{1}{M}T_{\mathrm{PN}} \xrightarrow[ M \rightarrow \infty]{} \begin{cases}
                            e^{-j \left(\sigma^{\phi}+\delta^{\varphi}\right)}~~~&\mathrm{CLO~setup}\\
                             e^{- \frac{\delta_{\phi}^{2}}{2}n-j \delta^{\varphi}}~~~&\mathrm{SLOs~setup}.                         \end{cases}
\end{align}
\end{lemma}
\proof The proof is straightforward by means of the application of the law of large numbers.
\endproof
\begin{remark}
Differently from \cite{Pitarokoilis2015} and  \cite{Bjornson2015}  that obtain the expectation of phase noise, this lemma provides the phase noise effect in the large number of antennas limit. Moreover, this lemma includes the effect of PN from both BS and UE LOs.
\end{remark}

Special focus has to be given to the transmit downlink signal $\bx \in \bbC ^{M \times 1}$ represented by~\eqref{transmit}. It is worthwhile to mention that under ideal hardware the  construction of the precoder employs the knowledge of current CSI,  which can be perfect or imperfect. In particular, under realistic conditions with RTHIs taken into account, where CSI is imperfect,  the sum-rate of the BC channel with uniform power allocation saturates at high SNR, if the channel error variance is fixed. In the ideal case of perfect CSI, the sum-rate remains unbounded as the SNR increases.

 For the sake of clarity and comparison, we present the conventional and RS performance metrics, namely the achievable sum-rates, below. 


\subsection{SINR with RTHIs and NoRS (Conventional Transmission)} \label{RS} 
After equal power allocation, we have  that the SINR  of UE $k$ is expressed by means of~\eqref{BasicSystemModelDownlink1}  as in~\eqref{nors}.
\begin{longequation*}[tp]
  \begin{align}
   \mathrm{SINR}_{k,n}^{\mathrm{NoRS}}=\frac{\frac{\rho_{k}}{K}{\lambda}|\bg_{k,n}^{\H}\bff_{k}|^2}{{\lambda}\sum_{j\ne k}^{K}\frac{\rho_{j}}{K}|\bg_{k,n}^{\H}\bff_{j}|^2+\EE\left[ |\bg_{k,n}^{\H}\etv_{\mathrm{t},n}^{\mathrm{BS}} |^2\right] +\EE\left[ |\eta_{\mathrm{r},n}^{\mathrm{UE}}|^2\right] +\xi_{k}^{\mathrm{UE}}}.\label{nors} 
\end{align}
\hrule
\end{longequation*}
Note that $\bg_{k,n}$ includes the channel evolution due to the effect of phase noise during the transmission phase. Furthermore, based on the worst-case assumption for the calculation of the mutual information~\cite[Lemma 1]{Bjornson2015}, we have treated the multi-user interference and distortion noises as independent Gaussian noises.

The  mutual information between the received signal and the transmitted symbols is lower bounded by the following achievable sum-rate. In particular, we have 
\begin{align}
 \mathrm{R}^{\mathrm{NoRS}}&=\sum_{k=1}^{K}\mathrm{R}_{k}^{\mathrm{NoRS}}\nn\\
 &=\frac{1}{T_{c}}\sum_{k=1}^{K}\sum_{n=1}^{T_{c}-\tau}\mathrm{R}_{k,n}^{\mathrm{NoRS}},\label{rate3} 
\end{align}
where $\mathrm{R}_{k,n}^{\mathrm{NoRS}}=\log_{2}\left( 1+ \mathrm{SINR}_{k,n}^{\mathrm{NoRS}} \right)$. Note that~\eqref{rate3} is obtained after following a similar approach to~\cite{Bjornson2015,Pitarokoilis2015}. In particular,  we compute the achievable rate of each UE for each time instance of the data transmission phase.

\subsection{SINR with RTHIs under RS}\label{RS} 
Given that our focus is to shed light on the behavior of the RS approach under RTHIs impairments, we apply uniform power allocation during the transmission of  the private messages for both conventional and RS cases. However, in the RS scenario, the powers allocated to the common and private parts are different\footnote{It is expected that RS will have a different impact in the additive RTHIs with comparison to the multiplicative RTHIs, since the former are power-dependent and will result in SINR degradation.}. Specifically, we allocate $\rho_{\mathrm{c}}=\rho\left( 1-t \right)$ to the common message and  $\rho_{ {k}}=\rho t/K$ to the private message of each UE, where $t \in \left( 0,1 \right]$. Actually, the role of $t$ is to adjust the  fraction of the total power spent for the transmission of the private messages.

According to this scheme, we have to evaluate the SINRs of both common and private messages. Since the transmit  signal is given by~\eqref{RStransmit}, these are given  by~\eqref{c1}-\eqref{c3}.
\begin{longequation*}
\begin{align}
   \mathrm{SINR}_{k,n}^{\mathrm{c}}&=\frac{{\rho_{\mathrm{c}}}{\lambda}|\bg_{k,n}^{\H} \bff_{c}|^2}{\lambda\sum_{j=1}^{K}{\frac{\rho_{j}}{K}|\bg_{k,n}^{\H} \bff_{j}|^2+\EE\left[ |\bg_{k,n}^{\H} \etv_{\mathrm{t},n}^{\mathrm{BS}} |^2\right] +\EE\left[ |\eta_{\mathrm{r},n}^{\mathrm{UE}}|^2\right] +\xi_{k}^{\mathrm{UE}}}}\label{c1} \\
  \mathrm{SINR}^{\mathrm{c}}_{n}&= \min_{k}\left( \mathrm{SINR}_{k,n}^{\mathrm{c}} \right)\label{c2} \\
  \mathrm{SINR}_{k}^{\mathrm{p}}&=\frac{\frac{\rho_{k}}{K}{\lambda}|\bg_{k,n}^{\H} \bff_{k}|^2}{{\lambda}\sum_{j\ne k}^{K}\frac{\rho_{j}}{K}|\bg_{k,n}^{\H} \bff_{j}|^2+\EE\left[ |\bg_{k,n}^{\H} \etv_{\mathrm{t},n}^{\mathrm{BS}} |^2\right] +\EE\left[ |\eta_{\mathrm{r},n}^{\mathrm{UE}}|^2\right] +\xi_{k}^{\mathrm{UE}}}\label{c3} .
  \end{align}
  \hrule
\end{longequation*}
   
The achievable sum-rate is written as 
\begin{align}
 \mathrm{R}^{\!\mathrm{RS}}=\mathrm{R}^{\mathrm{c}}+\sum_{j=1}^{K}\mathrm{R}_{j}^{\mathrm{p}},\label{RSSumRate} 
\end{align}
where, similar to~\eqref{rate3}, we have $\mathrm{R}^{\mathrm{c}}\!=\!\frac{1}{T_{c}}\!\!\sum_{n=1}^{T_{c}-\tau}\!\log_{2}\!\left( 1\!+\! \mathrm{SINR}_{n}^{\mathrm{c}} \right)$  and $\mathrm{R}_{j}^{\mathrm{p}}=\frac{1}{T_{c}}\sum_{n=1}^{T_{c}-\tau}\log_{2}\left( 1+ \mathrm{SINR}_{j,n}^{\mathrm{i}} \right)$  corresponding to the common and private achievable rates, respectively. Note that $\mathrm{SINR}_{n}^{\mathrm{c}} = \displaystyle  \min_{k}\left( \mathrm{SINR}_{k,n}^{\mathrm{c}} \right)$ and $\mathrm{SINR}_{j,n}^{\mathrm{p}}$ correspond to the common and private SINRs, respectively.
\section{Deterministic Equivalent Downlink Performance Analysis with RTHIs and Imperfect CSIT}\label{Deterministic} 
This section presents the precoder design for the common message, implemented to be used under the RS approach, and the main results corresponding to the DEs of the SINRs characterizing the transmissions of the common  and the private messages of UE $k$.

The DEs of the  SINRs for $\mathrm{NoRS}$ and $\mathrm{RS}$ are such that $\mathrm{SINR}_{k,n}-\overbar{\mathrm{SINR}}_{k,n}\xrightarrow[M \rightarrow \infty]{\mbox{a.s.}}0$\footnote{Note that $\xrightarrow[ M \rightarrow \infty]{\mbox{a.s.}}$ denotes almost sure convergence, and  $a_n\asymp b_n$ expresses the equivalence relation $a_n - b_n  \xrightarrow[ M \rightarrow \infty]{\mbox{a.s.}}  0$ with $a_n$  and $b_n$  being two infinite sequences.}, while the deterministic rate of UE $k$ is obtained by the dominated  convergence~\cite{Billingsley2008} and the continuous mapping theorem~\cite{Vaart2000} by means of~\eqref{rate3}, \eqref{RSSumRate} 
\begin{align}
R_{k}^{\mathrm{i}}-\bar{R}_{k}^{\mathrm{i}} \xrightarrow[ M \rightarrow \infty]{\mbox{a.s.}}0~~~\mathrm{i},=\mathrm{NoRS}, \mathrm{RS}\label{DeterministicSumrate}
\end{align}
where $\overbar{\mathrm{SINR}}_{k,n}$ and $\bar{R}_{k}$ are the corresponding DEs.
\subsection{Precoder Design}\label{PD} 
Under the assumption of linear precoding, the RS method entails two types of precoders multiplying the private and common messages, respectively. Regarding the design of the former, we note that in the case of a MISO BC with imperfect CSI the optimal precoder has to be optimized numerically~\cite{Dai2016}, while when perfect CSI is available, it takes the form of RZF~\cite{Bjornson2012Optimal}.  For the sake of simplicity, we design the  precoder of the private message by using RZF, as mentioned in the previous section. We elaborate further on this  below.
\subsubsection{Precoding of the Private Messages}
Due to the prohibitive complexity of large MIMO systems, we employ RZF for the transmission of the private messages constructed by means of the channel estimate $\hat{\bG}_{n}$. Thus, the BS  designs its RZF precoder as~\cite{Hoydis2013}
\begin{align}
\bF_{n}
&= \left(\tilde{\kappa}_{\mathrm{t}_\mathrm{BS}}^{2}\hat{\bW} \!+\!\kappa_{\mathrm{r}_\mathrm{UE}}^{2} \mathrm{diag}\left(\! \hat{\bW} \!\right)\!+\!  \bZ \!+\! M \al~\! \xi_{\mathrm{BS}} \Id_M\right)^{-1}{\hat{\bG}}\nn\\
&=  {\bSigma} {\hat{\bG}}, \label{eq:precoderRZF}
\end{align}
where we define
\begin{align}
 {\bSigma}&\triangleq\left(\tilde{\kappa}_{\mathrm{t}_\mathrm{BS}}^{2}\hat{\bW} \!+\!\kappa_{\mathrm{r}_\mathrm{UE}}^{2} \mathrm{diag}\left(\! \hat{\bW} \!\right)\!+\!  \bZ \!+\! M \al~\! \xi_{\mathrm{BS}} \Id_M\right)^{-1}
\end{align}
with $\tilde{\kappa}_{\mathrm{t}_\mathrm{BS}}^{2}\triangleq\left( 1+\kappa_{\mathrm{t}_\mathrm{BS}}^{2}\right)$ and $\hat{\bW}\triangleq\hat{\bG}\hat{\bG}^{\H}$.  Herein, $\bZ \in \bbC^{M \times M}$ is an arbitrary Hermitian
nonnegative definite matrix  and $\al$ is a regularization parameter scaled by $M$, in order to converge to a constant, as $M$, $K\to \infty$. Both  $\al$, $\bZ$ could be optimized, but this is outside the
scope of this paper and it is left for future work. 
\subsubsection{Precoding of the Common Message}
Following a similar procedure to~\cite{Dai2016}, we elaborate on the design of the precoder $\bff_{c}$ of the common message in the presence of RTHIs. Having in mind that in the large number of antennas regime the  different channel estimates tend to be orthogonal, we assume that $\bff_{c}$ can be written as a linear sum of these channel estimates in the subspace including $\hat{\bG}$, $\mathcal{S}=\mathrm{Span}\left( \hat{\bG} \right)$, i.e., it is given in the form of weighted matched beamforming. More concretely, we write
\begin{align}
 \bff_{c}=\sum_{k}\alpha_{k} \hat{\bg}_{k}.
\end{align}

The target is the maximization of the achievable
rate of the common message $\mathrm{R}_{k,n}^{\mathrm{c}}$. This optimization problem is described by
\begin{align}\begin{split}
&\mathcal{P}_{1}~:~\max_{\bff_{c} \in \mathcal{S}}\,\min_{k} {q}_{k}|\bg_{k,n}^{\H} \bff_{c}|^2,\\
 &\mathrm{s.t.}~~~~\|\bff_{c}\|^{2}=1\label{P1} 
\end{split}
\end{align}
where $q_{k}=\frac{{\rho_{\mathrm{c}}}\lambda}{\lambda\sum_{j=1}^{K}\frac{\rho_{j}}{K}|\bg_{k,n}^{\H} \bff_{j}|^2+\EE\left[ |\bg_{k,n}^{\H} \etv_{\mathrm{t},n}^{\mathrm{BS}} |^2\right] +\EE\left[ |\eta_{\mathrm{r},n}^{\mathrm{UE}}|^2\right] +\xi_{k}^{\mathrm{UE}}}$. The optimal solution $\{\al_{k}^{*}\}$ is provided by means of the following proposition.
Note that below, we are going to use the DE of $T_{\mathrm{PN}}$, given by  Lemma~\ref{SetupLOs}.

\begin{proposition}
In the large system limit, the optimal solution of the practical problem set by $\mathcal{P}_{1}$, where RTHIs are taken into account, is given by
\begin{align}
 \al^{*}_{k}=\frac{1}{\sqrt{M \sum_{j=1}^{K}\frac{q_{k}\frac{1}{M^{2}}\tr^{2}\hat{\bR}_{k}}{q_{j} \frac{1}{M^{2}}\tr^{2}\hat{\bR}_{j}}}},~\forall k.
\end{align}
\end{proposition}
\proof  Deriving the DEs of the equation and the constraint comprising the optimization problem described by $\mathcal{P}_{1}$, we lead to an optimization problem with deterministic variables. Specifically, applying~\eqref{eq:oneVector} from Lemma~\ref{lemma:asymptoticLimits} to \eqref{P1}, we obtain\footnote{Note that $\mathcal{P}_{2}$ includes a complex expression by means of $\widetilde\bTheta_{k,n}$.}
 \begin{align}\begin{split}
&\mathcal{P}_{2}~:~\max_{\al_{k} }\,\min_{k} {q}_{k}\frac{1}{M^{2}} |\al_{k}\tr \widetilde\bTheta_{k,n} \tr\hat{\bR}_{k}\
|^{2},\\
 &\mathrm{s.t.}~~~~\sum_{k}\al_{k}^{2}=\frac{1}{M}.\label{p2} 
\end{split}
\end{align}
Use of Lemmas~\ref{SetupLOs}, \ref{lemma:asymptoticproduct} transforms~\eqref{p2} to
\begin{align}
 &\mathcal{P}_{3}~:~~\max_{\al_{k} }\,\min_{k}  q_{k}\al_{k}^{2} \frac{1}{M^{2}}\begin{cases}
                          \tr^{2}\hat{\bR}_{k}~~~&\mathrm{CLO~setup}\\
                             e^{- {\sigma_{\phi}^{2}}n}\tr^{2}\hat{\bR}_{k}~~~&\mathrm{SLOs~setup},
                       \end{cases}\\
                             &\mathrm{s.t.}~~~~\sum_{k}\al_{k}^{2}=\frac{1}{M}.
\end{align}
Lemma $2$ in~\cite{Xiang2014} concludes the proof by enabling us to show that the optimal solution, satisfying $\mathcal{P}_{3}$, results, if   all terms are equal. In other words, when $q_{k}\al_{k}^{2} \frac{1}{M^{2}}\!\!\begin{cases}\!
                          \tr^{2}\hat{\bR}_{k}=q_{j}\al_{j}^{2} \frac{1}{M^{2}}\tr^{2}\hat{\bR}_{j}&\!\!\!\!\mathrm{CLO~setup}\\
                             \!e^{- {\sigma_{\phi}^{2}}n}\tr^{2}\hat{\bR}_{k}=q_{j}\al_{j}^{2} \frac{1}{M^{2}}e^{- {\sigma_{\phi}^{2}}n}\tr^{2}\hat{\bR}_{j}&\!\!\!\!\mathrm{SLOs~setup},
                       \end{cases}$ $\forall k\ne j$
\endproof

\subsection{Achievable Deterministic  Sum-Rate with RS in the Presence of RTHIs with Imperfect CSIT}
In this section, we present the DE analysis of a practical system with RTHIs during its data transmission,  which takes place for $T-\tau$ time slots. Actually, we conduct a DE analysis for both the RS and the NoRS strategies. Specifically, we derive the DE of the $k$th UE in the asymptotic limit of $K, M$ for fixed ratio $\beta=K/M$.

\begin{Theorem}\label{theorem:RZF}
The downlink DEs of the SINRs of UE $k$ at time $n$, corresponding to the  private and common messages with RZF precoding in the presence of RTHIs and imperfect CSIT, are given by~\eqref{privateSINR} and~\eqref{CommonSINR}
\begin{figure*}
\begin{align}
\overbar{\mathrm{SINR}}_{k}^{\mathrm{p}}&=\frac{\frac{\rho_{k}}{K}\bar{\lambda}\left( \frac{\frac{1}{M}\tr \widetilde\bTheta_{k,n}{\delta}_{k}}{1+{\delta}_{k} } \right)^{2}}{{\bar{\lambda}}\sum_{j\ne k}^{K}\frac{\rho t}{K}\frac{{Q}_{jk}}{M\left(1+{\delta_{j}}\right)^{2}}+ \frac{\rho t}{K} \kappa_{\mathrm{t}_\mathrm{BS}}^{2}  \frac{1}{M}\tr{\bR}_{k}+\frac{\rho t}{K} \kappa_{\mathrm{r}_\mathrm{UE}}^{2}\frac{1}{M}\tr{\bR}_{k}+\xi_{k}^{\mathrm{UE}}}\label{privateSINR} \\
\overbar{\mathrm{SINR}}_{k}^{c}&=\frac{{\rho_{\mathrm{c}}}\bar{\lambda}\left( \alpha_{k} \frac{1}{M}\tr\widetilde\bTheta_{k,n} \frac{1}{M}\tr \hat{\bR}_{k} \right)^2}{\bar{\lambda}{\frac{\rho t}{K}\left( \frac{\frac{1}{M}\tr \widetilde\bTheta_{k,n}{\delta}_{k}}{1+{\delta}_{k} } \right)^{2}+\sum_{j\ne k}^{K}\frac{\rho t}{K}\frac{{Q}_{jk}}{M\left(1+{\delta_{j}}\right)^{2}}+\frac{\rho t}{K} \kappa_{\mathrm{t}_\mathrm{BS}}^{2}  \frac{1}{M}\tr{\bR}_{k}+\frac{\rho t}{K} \kappa_{\mathrm{r}_\mathrm{UE}}^{2}\frac{1}{M}\tr{\bR}_{k}+\xi_{k}^{\mathrm{UE}}}}.\label{CommonSINR}
 \end{align}
 \line(1,0){470}
\end{figure*}
 where 
  \begin{align}
  \bar{\lambda}&=K\left( \frac{1}{M}\sum_{k=1}^{K}\frac{{\delta}_{k}^{'}}{\left( 1+{\delta}_{k} \right)^{2}} \right)^{-1},\nn
  \end{align}
  and
  \begin{align}
\!\!{Q}_{jk}\!\asymp\! \frac{ \delta_{j}^{''}}{M}\!\!+\!\frac{\left|{\delta_{k}^{''}}\right|^{2}\delta_{k}^{''}}{M\left( 1\!+\!\delta_{j} \right)^{2}}\!-\!2\mathrm{Re}\left\{ \! \frac{\frac{1}{M}\tr\widetilde\bTheta_{k,n} {\delta}_{k}\delta_{k}^{''} }{M\left( 1\!+\!\delta_{j} \right)}\!\right\}\!.
 \label{eq:theorem2.I.mu}
\end{align}
Also, we have 
${\delta}_{k}=\frac{1}{M}\tr \hat{\bR}_{k}\bT$, $\delta_{k}^{'}=\frac{1}{M}\tr \hat{\bR}_{k}\hat{\bT}^{'}$, $\delta_{j}^{'}=\frac{1}{M}\tr \hat{\bR}_{j}\hat{\bT}^{'}$, $\delta_{k}^{''}=\frac{1}{M}\tr \hat{\bR}_{k}\hat{\bT}^{''}$, ${\bS}=\left( \kappa_{\mathrm{r}_\mathrm{UE}}^{2} \mathrm{diag}\left(\! \hat{\bR}_{k} \!\right)+\bZ \right)/M$, and $\tilde{a}=\al  \xi_{\mathrm{BS}} $
 where
\begin{itemize}
\renewcommand{\labelitemi}{$\ast$}
\item $\bT=\bT(\tilde{a})$ and ${\deltav}=[{\delta}_{1},\cdots,{\delta}_{K}]^\T={\deltav}(\tilde{a})={\ev}(\tilde{a})$ are given by Theorem~\ref{th:detequ} for ${\bS}=\bS$, $\bD_k=\tilde{\kappa}_{\mathrm{r}_\mathrm{BS}}\hat{\bR}_{k}\, \forall k \in \mathcal{K}$,
\item  $\bT^{'}=\bT^{'}(\tilde{a})$ is given by Theorem~\ref{th:detequder} for  ${\bS}=\bS$, $\bK=\Id_M$, $\bD_k=\tilde{\kappa}_{\mathrm{r}_\mathrm{BS}}\hat{\bR}_{k}, \forall  k \in \mathcal{K}$,
\item  $\bT^{''}=\bT^{''}(\tilde{a})$ is given by Theorem~\ref{th:detequder} for ${\bS}=\bS$, $\bK= \hat{\bR}_{k}$, $\bD_k=\tilde{\kappa}_{\mathrm{r}_\mathrm{BS}}\hat{\bR}_{k}, \forall  k \in \mathcal{K}$.
\end{itemize} 
\end{Theorem}
\proof The proof of Theorem~\ref{theorem:RZF} is given in Appendix~\ref{theorem3}.\endproof
\begin{remark}
Clearly, the various impairments decrease the SINR of  both common and private messages. Especially, given that the additive distortions and amplified thermal noise are found in the denominator of the SINRs, their increase degrades the system performance. Hence, the rate saturates at high SNR. Nevertheless, the PN affects both the numerator and the denominator of the SINRs by putting an extra penalty to the quality of the system. Its increase results in an obvious decrease of the numerator, and an increase of the denominator by means of an increase of ${Q}_{jk}$.
\end{remark}
\subsection{Power Allocation}
Given that the optimal $t$ would be rather intricate if it was obtained by maximizing~\eqref{RSSumRate} after calculating its first derivative, this method is not indicated. For this reason, we turn our attention to find a suboptimal, but effective solution by following the example in~\cite{Dai2016}. Specifically, the target is to fulfil the condition, which allows RS  to outperform the conventional multi-user broadcasting. Achievement of this condition can be accomplished by allocating  a fraction $t$ of the
total power for the  transmission of the private messages  RS, in order to realize almost the same  sum-rate as the conventional 
BC with full power. Exploitation of the remaining power to transmit the common message enables   RS to boost the sum-rate at high-SNR. The sum-rate payoff of the RS strategy over the conventional BC (NoRS) can be determined by the difference
\begin{align}
 \Delta R=\mathrm{
 R}^{\mathrm{c}}+\sum_{k=1}^{K}\left( \mathrm{R}_{k}^{\mathrm{p}}-\mathrm{R}_{k}^{\mathrm{NoRS}} \right).
\end{align}

The necessary condition and the power
splitting ratio $t$ are given by the following proposition, which enables RS  outperform conventional multiuser broadcasting.
\begin{proposition}\label{prop:inequality} 
We can write 
\begin{align}
 \mathrm{R}_{k}^{\mathrm{p}}\le 
\mathrm{R}_{k}^{\mathrm{NoRS}}.
\label{inequality} 
\end{align}
The equality holds when the power splitting ratio $t$ is given by
  \begin{align}
\!\!\!t=\min\bigg\{\frac{K^{2}M }{{\bar{\lambda}}\!\sum_{j\ne k}^{K}\!\frac{\rho {Q}_{jk}}{\left(1+{\delta_{j}}\right)^{2}\!}\!+\!\bar{m}_{k}\rho \left(   \kappa_{\mathrm{t}_\mathrm{BS}}^{2}\!+\! \kappa_{\mathrm{r}_\mathrm{UE}}^{2} \right)+M \xi_{k}^{\mathrm{UE}}},1\bigg\},\label{tau} 
\end{align}
where $\bar{m}_{k}=\tr \hat{\bR}_{k}$.
In such case, the sum-rate gain $\Delta R$ becomes
\begin{align}
 \Delta R\ge \mathrm{R}^{\mathrm{c}}-\log_2 e.
\end{align}
\end{proposition}
\proof See Appendix~\ref{proofinequality}.\endproof
\begin{remark}
Interestingly, \eqref{tau} allows us to make insightful observations regarding the power allocation and its impact on the sum-rate. The dependence of  the system parameters and the RTHIs on $t$  is noteworthy. Specifically, increasing the severity of any of the RTHIs results in less power allocated to the private messages. Moreover, at high SNR $\rho t$ becomes independent of $\rho$, while the sum-rate
increases with the available transmit power by assigning the
remaining power $\rho-\rho t$ to the common message. On the contrary at low SNR, $t=1$, which means that the common message becomes useless. In other words, RS degenerates to NoRS, where broadcasting of only  private messages takes place.  Generally, by increasing the RTHIs, RS presents more its robustness. 
\end{remark}

\section{Deterministic Equivalent Downlink Performance Analysis with RTHIs and Perfect CSIT}\label{DeterministicPerfect}
This section provides the design of the MISO BC with RTHIs in terms of the  asymptotic (DEs) SINRs, but we omit the proofs, since these are straightforwardly established after following the same analysis with Section~\ref{Deterministic}. The purpose of this section is to provide the means for comparison (benchmark) with the corresponding expressions regarding imperfect CSIT.
\subsection{Precoding Design}
The BS  designs the RZF precoder of the private messages  as
\begin{align}
\bF_{n}
&= \left(\tilde{\kappa}_{\mathrm{r}_\mathrm{BS}}^{2}{\bW} \!+\!\kappa_{\mathrm{r}_\mathrm{UE}}^{2} \mathrm{diag}\left(\! {\bW} \!\right)\!+\!  \bZ \!+\! M \al~\! \xi_{k}^{\mathrm{UE}} \Id_M\right)^{-1}{{\bG}}\nn\\
&=  {\bSigma} {{\bG}}, \label{eq:precoderRZF}
\end{align}
where we define
\begin{align}
{\bSigma}&\triangleq\left(\tilde{\kappa}_{\mathrm{r}_\mathrm{BS}}^{2}{\bW} \!+\!\kappa_{\mathrm{r}_\mathrm{UE}}^{2} \mathrm{diag}\left(\! \hat{\bW} \!\right)\!+\!  \bZ \!+\! M \al~\! \xi_{k}^{\mathrm{UE}} \Id_M\right)^{-1}
\end{align}
with ${\bW}\triangleq{\bG}{\bG}^{\H}$ and the rest of parameters as in Subsection~\ref{PD}.

As far as the precoding of the common message is concerned, it is given by
\begin{align}
 \bff_{c}=\sum_{k}\alpha_{k} {\bg}_{k},
\end{align}
where 
\begin{align}
 \al^{*}_{k}=\frac{1}{\sqrt{M \sum_{j=1}^{K}\frac{q_{k}\frac{1}{M^{2}}\tr^{2}{\bR}_{k}}{q_{j} \frac{1}{M^{2}}\tr^{2}{\bR}_{j}}}},~\forall k
\end{align}
with $q_{k}$ defined below~\eqref{P1}.

\subsection{Achievable Deterministic  Sum-Rate with RS in the Presence of RTHIs with Perfect CSIT}
In this section, we present the DE analysis of a practical system with RTHIs during its data transmission but with perfect CSIT,  which takes place for $T$ time slots. Its implementation is based on the RS method. Specifically, we derive the DE of the $k$th UE in the asymptotic limit of $K, M$ for fixed ratio $\beta=K/M$, when RZF is employed.

\begin{Theorem}\label{theorem:RZF1}
The downlink DEs of the SINRs of UE $k$ at time $n$, corresponding to the  private and common messages with RZF precoding in the presence of RTHIs  and perfect CSIT, are given by~\eqref{privateSINRPer} and~\eqref{CommonSINRPer}
\begin{figure*}
\begin{align}
\overbar{\mathrm{SINR}}_{k}^{\mathrm{p}}&=\frac{\frac{p_{k}}{K}\bar{\lambda}\left( \frac{\frac{1}{M}\tr \widetilde\bTheta_{k,n}{\delta}_{k}}{1+{\delta}_{k} } \right)^{2}}{{\bar{\lambda}}\sum_{j\ne k}^{K}\frac{pt}{K}\frac{{Q}_{jk}}{M\left(1+{\delta_{j}}\right)^{2}}+ \frac{pt}{K} \kappa_{\mathrm{t}_\mathrm{BS}}^{2}  \frac{1}{M}\tr\ {\bR}_{k}+\frac{pt}{K} \kappa_{\mathrm{r}_\mathrm{UE}}^{2}\frac{1}{M}\tr\ {\bR}_{k}+\xi_{k}^{\mathrm{UE}}}\label{privateSINRPer} \\
 \overbar{\mathrm{SINR}}_{k}^{c}&=\frac{{p_{\mathrm{c}}}\bar{\lambda}\left( \alpha_{k} \frac{1}{M}\tr\widetilde\bTheta_{k,n} \frac{1}{M}\tr \ {\bR}_{k} \right)^2}{\bar{\lambda}{\frac{p t}{K}\left( \frac{\frac{1}{M}\tr \widetilde\bTheta_{k,n}{\delta}_{k}}{1+{\delta}_{k} } \right)^{2}+\sum_{j\ne k}^{K}\frac{p t}{K}\frac{{Q}_{jk}}{M\left(1+{\delta_{j}}\right)^{2}}+\frac{pt}{K} \kappa_{\mathrm{t}_\mathrm{BS}}^{2}  \frac{1}{M}\tr\ {\bR}_{k}+\frac{pt}{K} \kappa_{\mathrm{r}_\mathrm{UE}}^{2}\frac{1}{M}\tr\ {\bR}_{k}+\xi_{k}^{\mathrm{UE}}}}.\label{CommonSINRPer} 
\end{align}
 \line(1,0){470}
\end{figure*}
 where 
  \begin{align}
  \bar{\lambda}&=K\left( \frac{1}{M}\sum_{k=1}^{K}\frac{{\delta}_{k}^{'}}{\left( 1+{\delta}_{k} \right)^{2}} \right)^{-1},\nn
  \end{align}
  and
  \begin{align}
\!\!{Q}_{jk}\!\asymp\! \frac{ \delta_{j}^{''}}{M}\!\!+\!\frac{\left|{\delta_{k}^{''}}\right|^{2}\delta_{k}^{''}}{M\left( 1\!+\!\delta_{j} \right)^{2}}\!-\!2\mathrm{Re}\left\{ \! \frac{\frac{1}{M}\tr\widetilde\bTheta_{k,n}{\delta}_{k}\delta_{k}^{''} }{M\left( 1\!+\!\delta_{j} \right)}\!\right\}\!.
 \label{eq:theorem2.I.mu}
\end{align}
 Also, we have
${\delta}_{k}\!=\!\frac{1}{M}\!\tr\! \ {\bR}_{k}\bT$, $\delta_{k}^{'}\!=\!\frac{1}{M}\!\tr\! \ {\bR}_{k}\ {\bT}^{'}$, $\delta_{j}^{'}\!=\!\frac{1}{M}\tr \ {\bR}_{j}\ {\bT}^{'}$, $\delta_{k}^{''}=\frac{1}{M}\tr \ {\bR}_{k}\ {\bT}^{''}$, ${\bS}=\left( \kappa_{\mathrm{r}_\mathrm{UE}}^{2} \mathrm{diag}\left(\! \ {\bR}_{k} \!\right)+\bZ \right)/M$, and $\tilde{a}=a \xi_{k}^{\mathrm{BS}} $
 where
\begin{itemize}
\renewcommand{\labelitemi}{$\ast$}
\item $\bT=\bT(\tilde{a})$ and ${\deltav}=[{\delta}_{1},\cdots,{\delta}_{K}]^\T={\deltav}(\tilde{a})={\ev}(\tilde{a})$ are given by Theorem~\ref{th:detequ} for ${\bS}=\bS$, $\bD_k=\tilde{\kappa}_{\mathrm{r}_\mathrm{BS}}\ {\bR}_{k}\, \forall k \in \mathcal{K}$,
\item  $\bT^{'}=\bT^{'}(\tilde{a})$ is given by Theorem~\ref{th:detequder} for  ${\bS}=\bS$, $\bK=\Id_M$, $\bD_k=\tilde{\kappa}_{\mathrm{r}_\mathrm{BS}}\ {\bR}_{k}, \forall  k \in \mathcal{K}$,
\item  $\bT^{''}=\bT^{''}(\tilde{a})$ is given by Theorem~\ref{th:detequder} for ${\bS}=\bS$, $\bK= \bR_{k}$, $\bD_k=\tilde{\kappa}_{\mathrm{r}_\mathrm{BS}}\ {\bR}_{k}, \forall  k \in \mathcal{K}$.
\end{itemize} 
\end{Theorem}

Similarly, the power splitting ratio $t$ during the power allocation of the scenario, where perfect CSIT is available, becomes
\begin{align}
 t=\min\bigg\{\frac{K^{2}M }{{\bar{\lambda}}\!\sum_{j\ne k}^{K}\!\frac{\rho {Q}_{jk}}{\left(1+{\delta_{j}}\right)^{2}\!}\!+\!\bar{m}_{k}\rho \left(   \kappa_{\mathrm{t}_\mathrm{BS}}^{2}\!+\! \kappa_{\mathrm{r}_\mathrm{UE}}^{2} \right)+M \xi_{k}^{\mathrm{UE}}}\bigg\},
\end{align}
where now  $\bar{m}_{k}=\tr {\bR}_{k}$.

\section{Numerical Results}\label{NumericalResults} 
This section presents the numerical illustrations of the analytical and Monte Carlo simulation results obtained for both cases of perfect and imperfect CSIT. Specifically, the bullets represent the simulation results. The black line depicts the ideal sum-rate with RZF and no common message (NoRS), i.e., perfect CSIT and hardware are assumed. The    blue and green lines depict the sum-rate with RS, when perfect and imperfect CSIT is considered.  respectively. For comparison, red and cyan   show the sum-rate with NoRS   when perfect  and  imperfect CSIT is assumed. The discrimination between ``solid'' and ``dot'' lines, where applicable, designates the  results with SLOs and CLO, respectively.

\subsection{Simulation Setup}
The simulation setting considers a cell  of $250~\mathrm{m}\times 250~\mathrm{m}$ with $K = 2$ UEs, where the   randomly selected UE is found at a distance of 25m from the BS. The pilot length is $B = 2$. Moreover, we assume a Rayleigh block-fading channel by  taking into consideration that the coherence time and the coherence bandwidth are $T_{c}=5~\mathrm{ms}$ and $B_{c}=100~\mathrm{KHz}$, respectively. Hence, the coherence block consists of $T = 500$ channel uses.  In each block, we assume fast fading by means of $\bw_{k} \sim \cC\cN(\b0,\Id_{M})$. Also, we set $\bR_{k}=\bm \Lambda_{k}$, i.e., we account for path-loss and shadowing, where $\Lambda_{k}$ is a $M\times M$ diagonal matrix with elements across the diagonal modeled as~\cite{Bjornson2015}
\begin{align}
 \lambda_{k}^{m}=\frac{10^{s_{k}^{m}-1.53}}{\left( d_{k}^{m} \right)^{3.76}},
\end{align}
where $d_{k}^{m}$ is the distance in meters between the  receive antenna $m$ and UE $k$, while $s_{k}^{m}\sim \cN\left(0,3.16\right)$ represents the shadowing effect. 

The  power of the uplink training symbols is  $\rho_{up}^{\mathrm{UE}}=2~\mathrm{dB}$ and the variance of thermal noise is assumed $\sigma^{2}= -174~ \mathrm{dBm}/\mathrm{Hz}$. Also, the PN is simulated as a discrete Wiener process with specific increment variance, and for the sake of simulations, we have set that the nominal values of the uplink RTHIs equal to the downlink RTHIs, e.g., $\xi^{\mathrm{BS}}=\xi^{\mathrm{UE}}$. 

\subsection{Impact of Hardware Impairments on NoRS/RS-Comparisons}
In the following figures, both perfect and imperfect CSIT scenarios are investigated. The metric under investigation is the DE sum-rate in the cases of  both NoRS  and RS strategies. The theoretical curves for the cases with imperfect and perfect CSIT are obtained by means of Theorems~\ref{theorem:RZF} and~\ref{theorem:RZF1}, while  the simulated curves are obtained by averaging the corresponding SINRs over $10^3$ random channel instances. Clearly, as can be shown from the figures,  although the DEs are   derived for $M$, $K\to \infty$ with a given ratio, the corresponding results concur with simulations for  finite values of $M$, $K$\footnote{Herein and without loss of generality, we consider the overall impact of the PN. Specifically, we add up both PN contributions coming from   BS CLO/SLOs and UE LO. In the case of additive impairments, we assume that $\kappa_{\mathrm{t}_\mathrm{BS}}^{2}=\kappa_{\mathrm{r}_\mathrm{UE}}^{2}=\kappa^{2}$.}. Note that $t$ used in the simulation is  obtained  by means of  both  exhaustive search and Proposition~\ref{prop:inequality} for verification.

\subsubsection{Multiplicative distortions (PN)}
\begin{figure}[!h]
 \begin{center}
 \includegraphics[width=\linewidth]{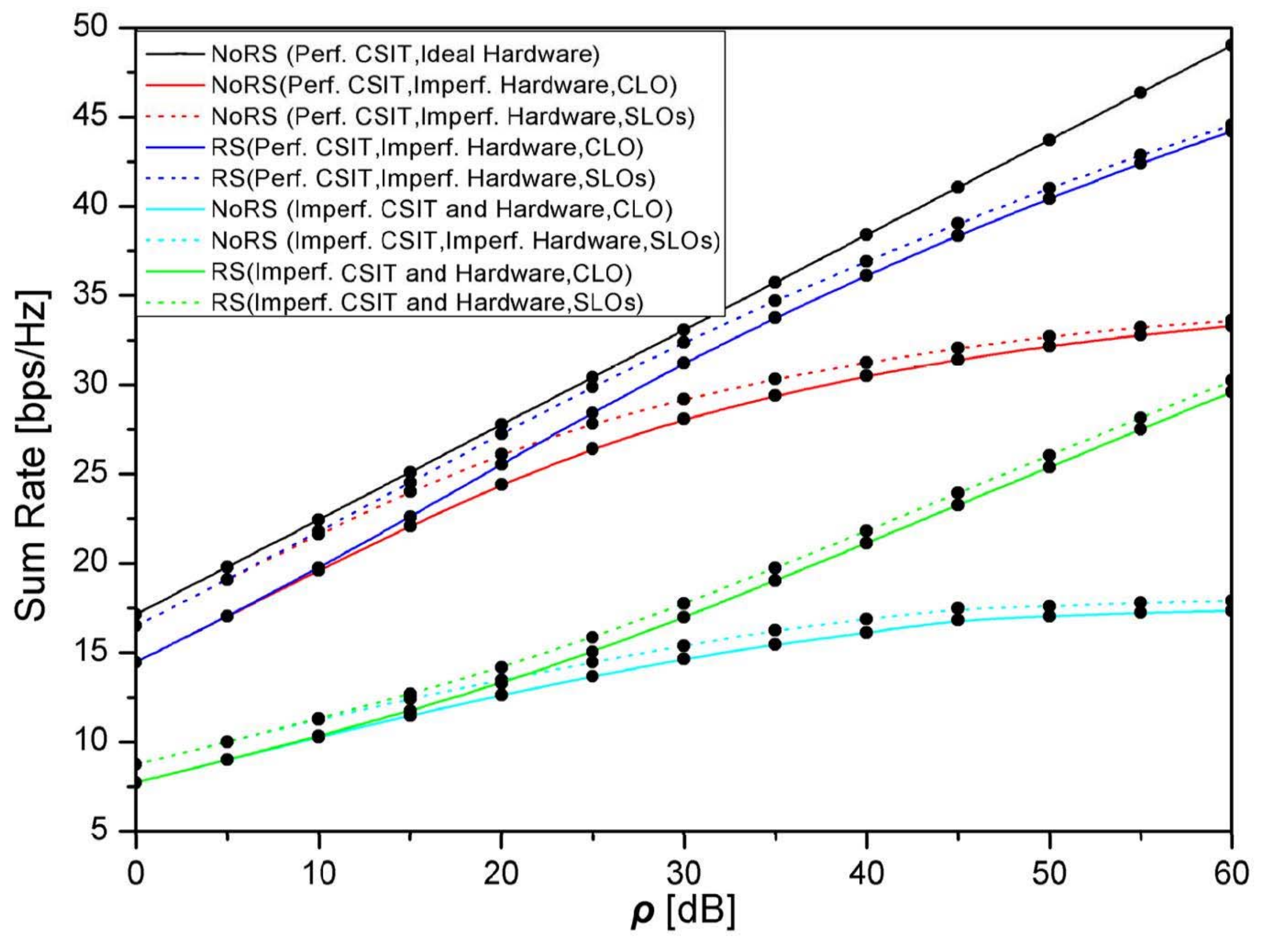}
 \caption{\footnotesize{Sum-rate versus $\rho$  ($M=100$, $K=2$, $T=500$, $\delta=10^{-4}$, $\kappa^{2}=0$, $\xi_{k}^{\mathrm{UE}}=\sigma^{2}$).}}
 \label{M=100}
 \end{center}
 \end{figure}

Fig.~\ref{M=100} provides the comparison of the sum-rate  versus the SNR in both cases of perfect and imperfect CSIT by considering  $M=100$ and $T=500$, while only  the total PN is taken into account on both the uplink and the downlink, i.e., we assume no additive impairments and amplified thermal noise.  The sum-rate with NoRS under perfect CSI and ideal hardware increases monotonically with the increase in the value of $\rho$. In the practical case where the  PN is considered, NoRS saturates after a certain value of SNR in cases of perfect (no RTHIs and no imperfect CSIT at the uplink stage) and imperfect CSIT. Remarkably, RS proves to be robust, since the sum-rate does not saturate. Moreover, when imperfect CSIT is assumed, the degradation of the sum-rate in all case is obvious. Notably, the setting with SLOs behaves better than the BS architecture with CLO because in such case the phase drifts are independent and in the large system limit they are averaged~\cite{Bjornson2015}. Of course, the employment of many LOs (SLOs) results in higher deployment cost. In the presence of PN, RS mitigates the multi-user interference due to imperfect CSIT. Thus, we have no saturation. The same effect occurs, when we have perfect CSIT, but the PN is present.
 
\begin{figure}[!h]
 \begin{center}
 \includegraphics[width=\linewidth]{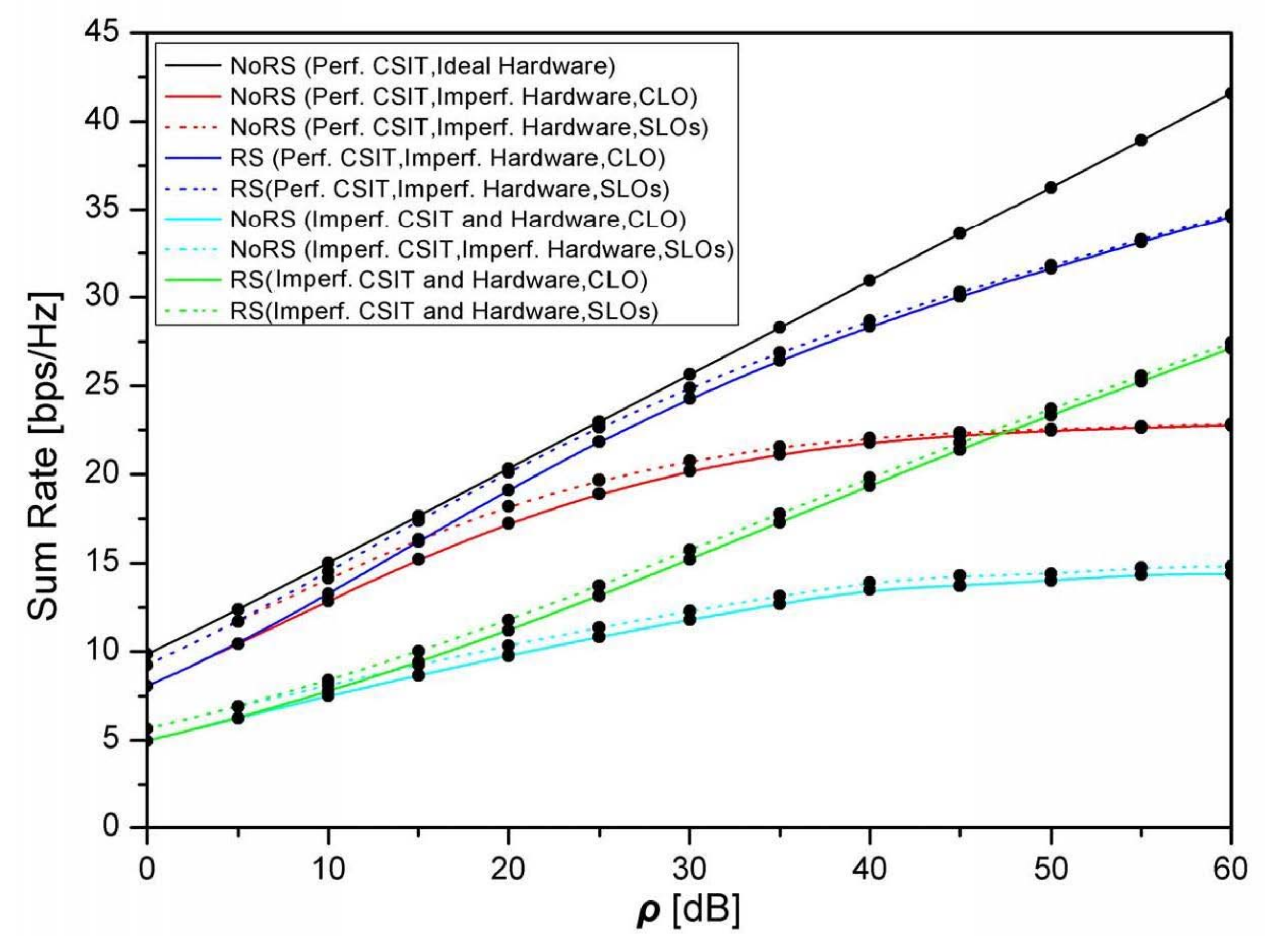}
 \caption{\footnotesize{Sum-rate versus $\rho$  ($M=20$, $K=2$, $T=500$, $\delta=10^{-4}$, $\kappa^{2}=0$, $\xi_{k}^{\mathrm{UE}}=\sigma^{2}$).}}
 \label{M=20}
 \end{center}
 \end{figure}

\begin{figure}[!h]
 \begin{center}
 \includegraphics[width=\linewidth]{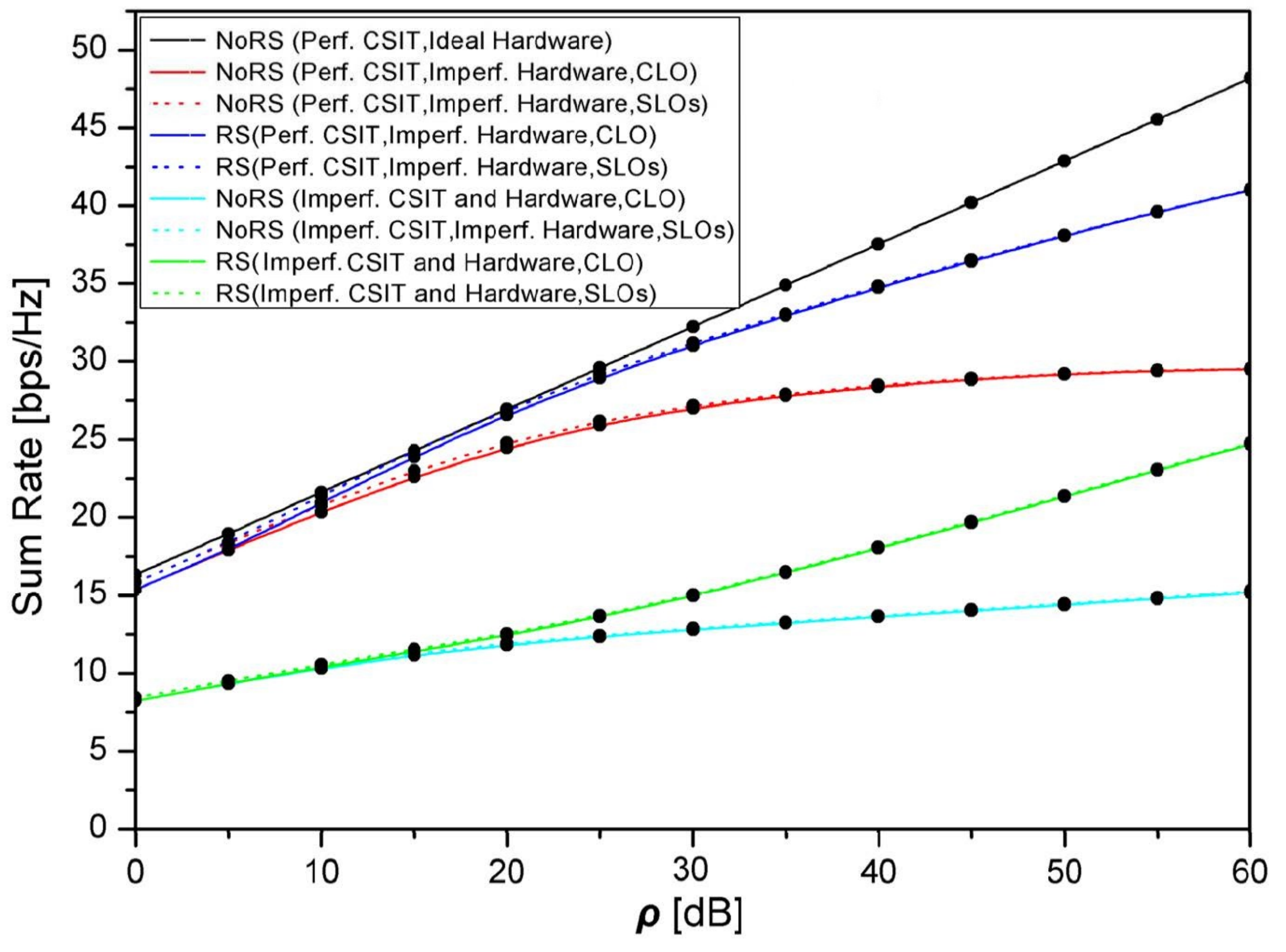}
 \caption{\footnotesize{Sum-rate versus $\rho$  ($M=100$, $K=2$, $T=100$, $\delta=10^{-4}$, $\kappa^{2}=0$, $\xi_{k}^{\mathrm{UE}}=\sigma^{2}$).}}
 \label{T=100}
 \end{center}
 \end{figure}

Fig.~\ref{M=20} presents the comparison of the  sum-rate versus the  SNR after decreasing the number of BS antennas to $M/5=20$, while Fig.~\ref{T=100} shows the impact of the coherence time $T$, when it is decreased to $T/5=100$ channel uses. In other words, in both cases we have an equivalent decrease of 1/5 of $M$ and $T$. From the former figure, we conclude a decrease of the sum-rate due to the corresponding decrease of $M$, as expected. Furthermore, the difference between the SLOs and CLO setups now is smaller because we have less independent LOs. The latter figure exposes that a decrease regarding the number of channel uses results in a small decrease of the sum-rate with the saturation in the case of NoRS taking place earlier. Also, the gap between the sum-rates with CLO and SLOs setups is smaller now.
 
\begin{figure}
\begin{center}
\subfigure[]{
\includegraphics[width=\linewidth]{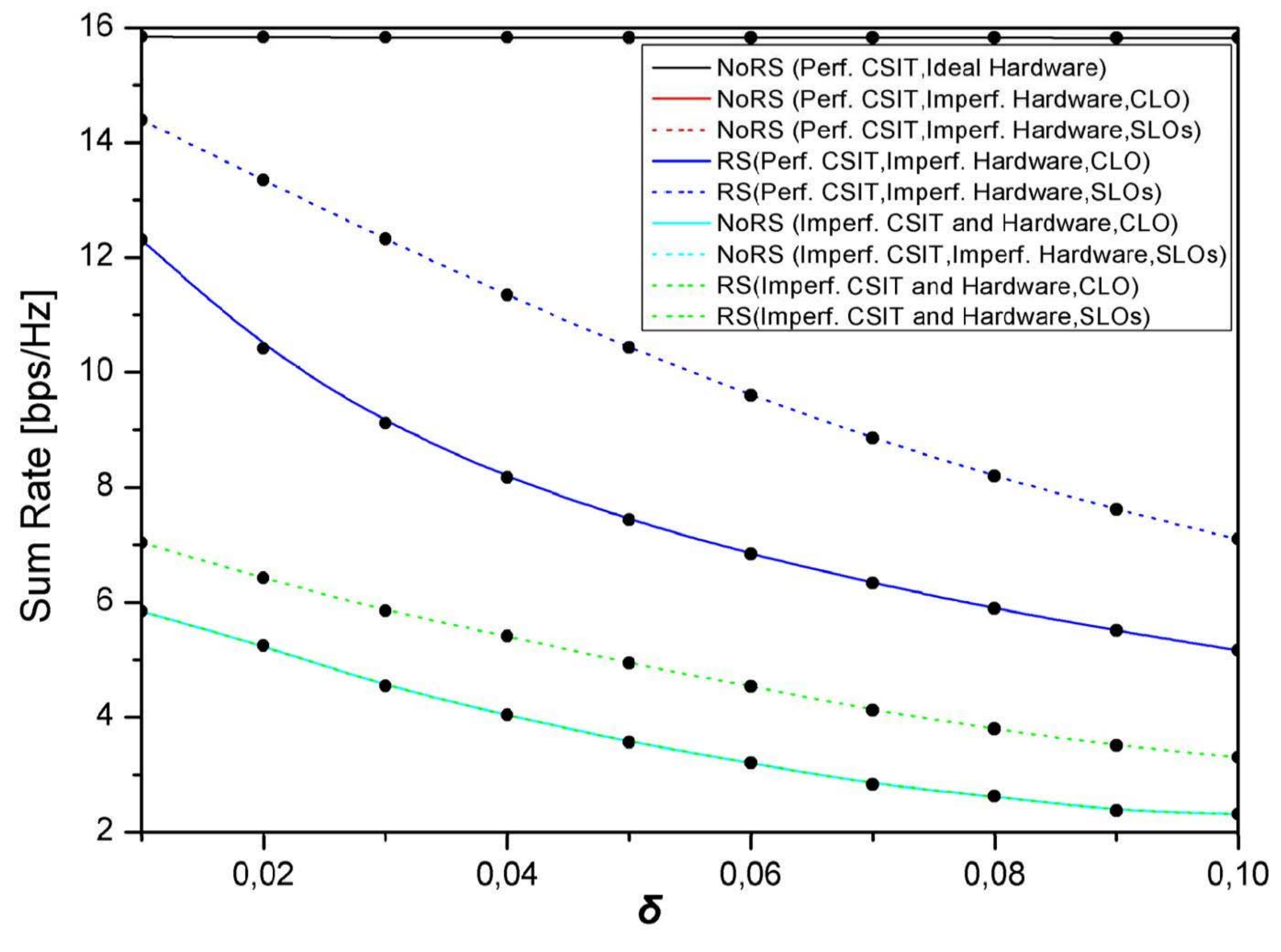}\label{25varyingdelta}
} \subfigure[]{
\includegraphics[width=\linewidth]{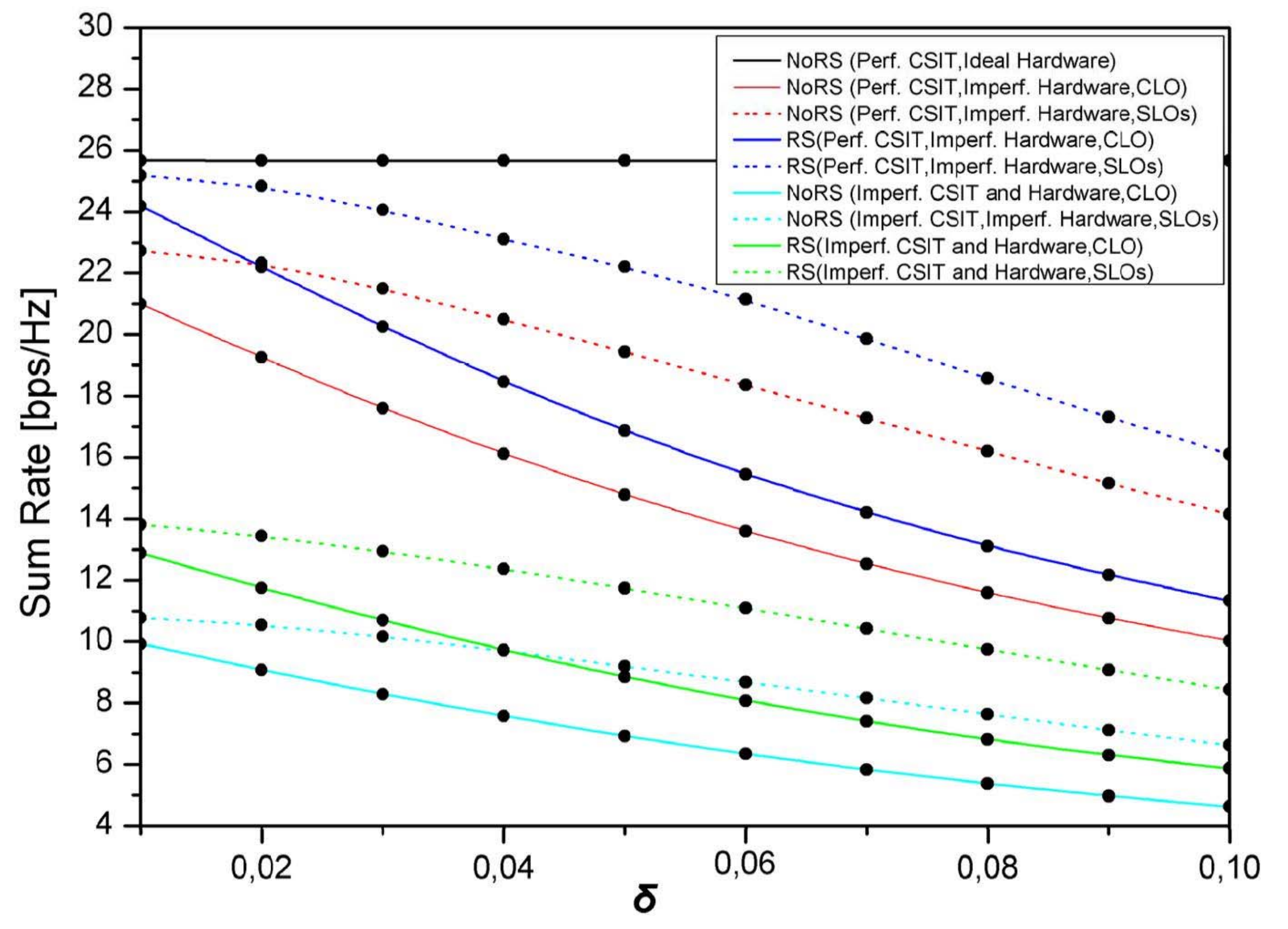}
\label{5varyingdelta}
}
\end{center}
\caption{\footnotesize{(a) Sum-rate versus $\delta$  ($M=100$, $K=2$, $T=500$, $\rho=5~\mathrm{dB}$, $\kappa^{2}=0$, $\xi=\sigma^{2}$). (b) Sum-rate versus $\delta$  ($M=100$, $K=2$, $T=500$, $\rho=25~\mathrm{dB}$, $\kappa^{2}=0$, $\xi_{k}^{\mathrm{UE}}=\sigma^{2}$).}}

\vspace{-10 pt}
\end{figure}

Figs.~\ref{25varyingdelta} and~\ref{5varyingdelta} illustrate the sum-rate versus the total PN coming from the BS CLO/SLOs and the LOs of the UEs for $\rho=5~\mathrm{dB}$ amd $\rho=25~\mathrm{dB}$, respectively. According to both figures, it can be noted that the various sum-rates  decrease monotonically with $\delta$, however, in the first figure, where the SNR is small, there is no improvement coming from the implementation of RS. Hence, the NoRS lines coincide with the respective lines corresponding to the RS strategy. Obviously, the expected improvement appears in the second figure, where a higher SNR is assumed. This is reasonable, since RS exhibits its outperformance at the high-SNR regime. Moreover, as $\delta$ decreases, the gap between the sum-rates corresponding to the SLOs and CLO setups narrows because the degradation coming from the accumulation of PN decreases.
\subsubsection{Additive distortions}
\begin{figure}
\begin{center}
\subfigure[]{
\includegraphics[width=\linewidth]{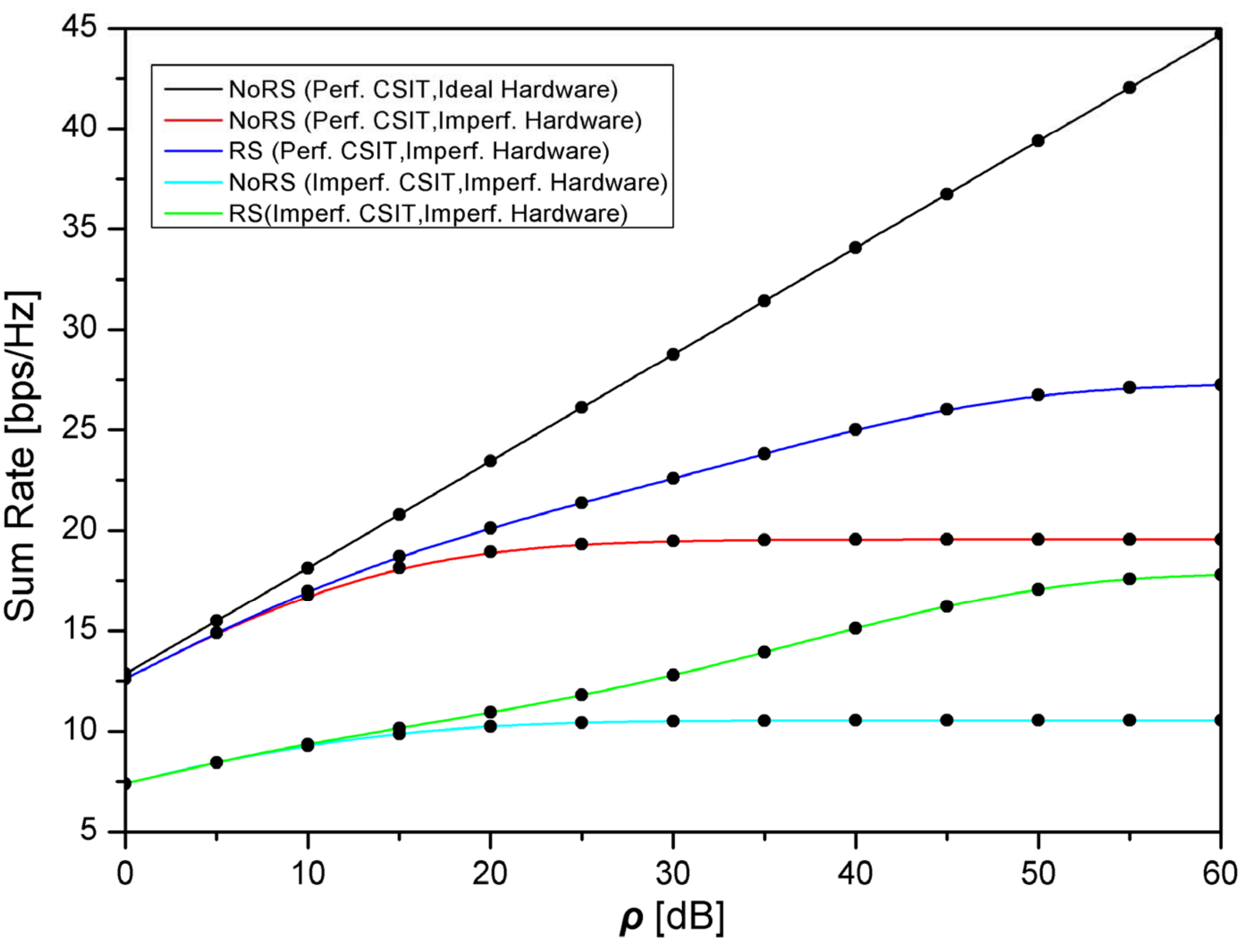}\label{00000156}
} \subfigure[]{
\includegraphics[width=\linewidth]{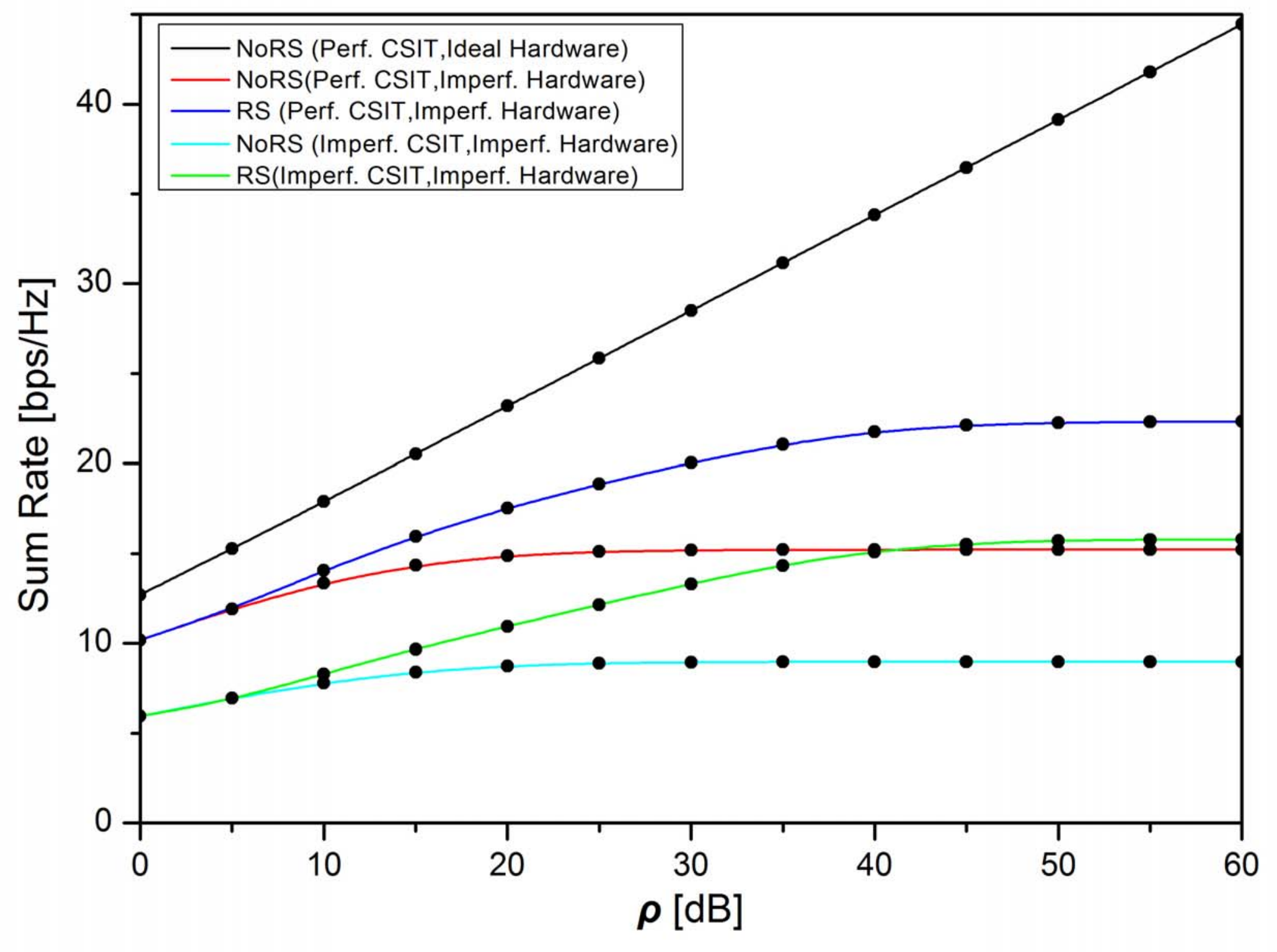}
\label{000156}
}
\end{center}
\caption{\footnotesize{(a) Sum-rate versus $\rho$  ($M=100$, $K=2$, $T=500$, $\delta=0$, $\kappa^{2}=1.56\cdot10^{-4}$, $\xi_{k}^{\mathrm{UE}}=\sigma^{2}$). (b) Sum-rate versus $\rho$  ($M=100$, $K=2$, $T=500$, $\delta=0$, $\kappa^{2}=1.56\cdot10^{-2}$, $\xi_{k}^{\mathrm{UE}}=\sigma^{2}$).}}

\vspace{-10 pt}
\end{figure}

Fig. \ref{00000156} presents the comparison among the various sum-rates versus $\rho$, when only additive RTHIs are considered, i.e., without any phase and amplified thermal noises. From the figure, it can be observed that RS is not robust in such case because it mitigates the multi-user interference due to imperfect CSIT, but it is not able to change the power dependence of the additive RTHIs. Interestingly, both sum-rates with NoRS and RS saturate at high SNR, even if we have available perfect CSIT. As mentioned, the reason behind this inadequacy of RS is hidden behind the power dependence of the additive RTHIs. Thus, power-dependent terms appearing in the denominator of the SINR lead to a noise-limited scenario and ultimately a saturation of the sum-rates. In Fig.~\ref{000156}, it is shown the further degradation of the sum-rates after increasing the impact of additive RTHIs to $\kappa^{2}=0.00156$.
\subsubsection{Amplified thermal  noise}
\begin{figure}[!h]
 \begin{center}
 \includegraphics[width=\linewidth]{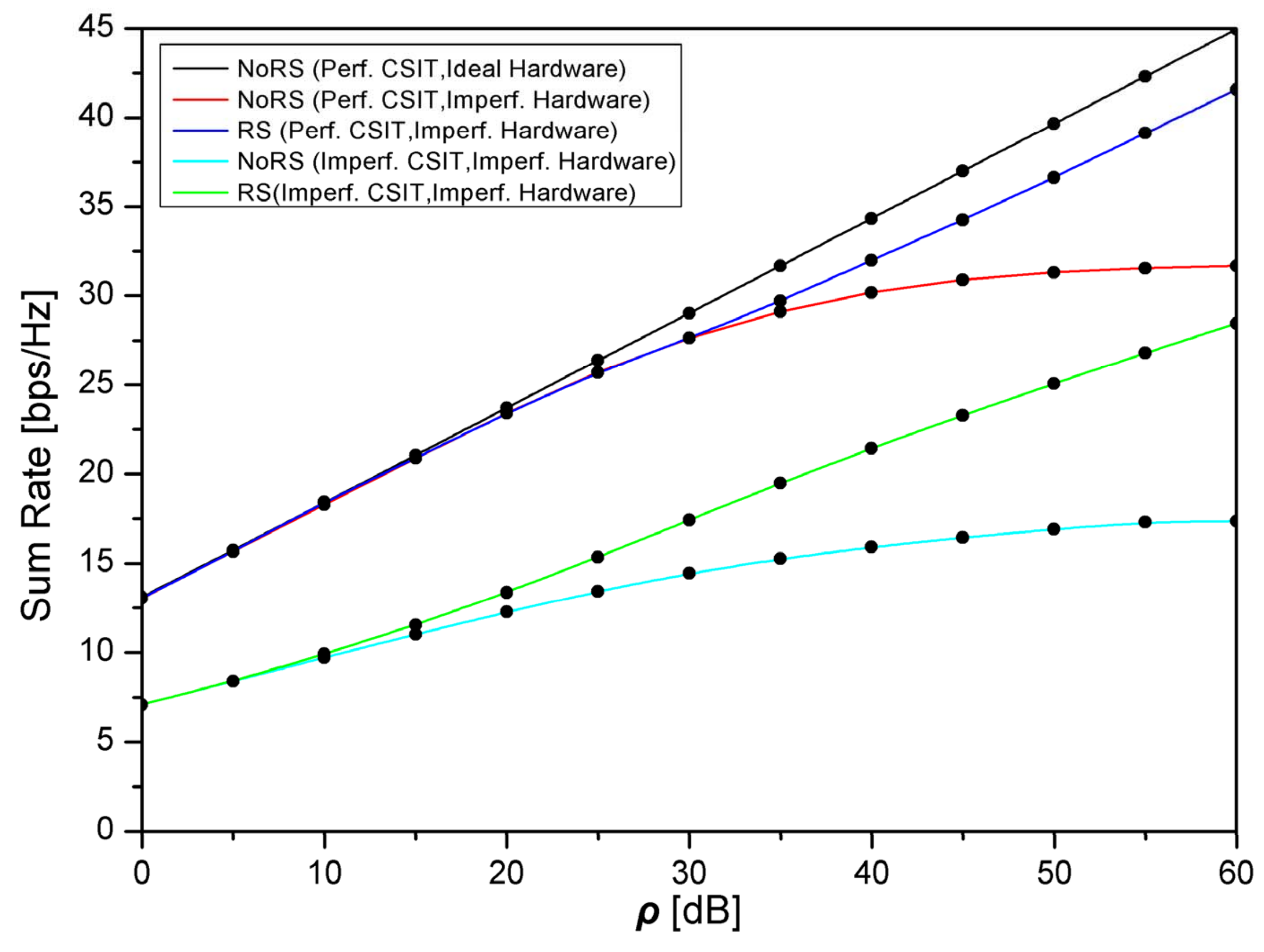}
 \caption{\footnotesize{Sum-rate versus $\rho$  ($M=100$, $K=2$, $T=500$, $\delta=0$, $\kappa^{2}=0$, $\xi_{k}^{\mathrm{UE}}=2\sigma^{2}$).}}
 \label{xi}
 \end{center}
 \end{figure}

In order to illustrate the effect of amplified thermal noise on the sum-rate, in Fig. \ref{xi}, we plot the latter versus $\rho$ for an increased value of $\xi_{k}^{\mathrm{UE}}$ equal to $2 \sigma^{2}$. Note that the rest of impairments, i.e., the PN and the additive ones are assumed that have no impact because our focus is to shed light on the impact of this separate effect. Thus, for this evaluation,  all other impairments values are considered to be zero while analyzing the effect of this impairment. From the figure, it can be noted that the sum-rates decrease with the increase in the value of $\xi_{k,n}^{\mathrm{UE}}$ for all cases.  Another interesting observation from our analysis is that RS keeps being robust in this case too, as in the case of PN.
 \begin{figure}[!h]
 \begin{center}
 \includegraphics[width=\linewidth]{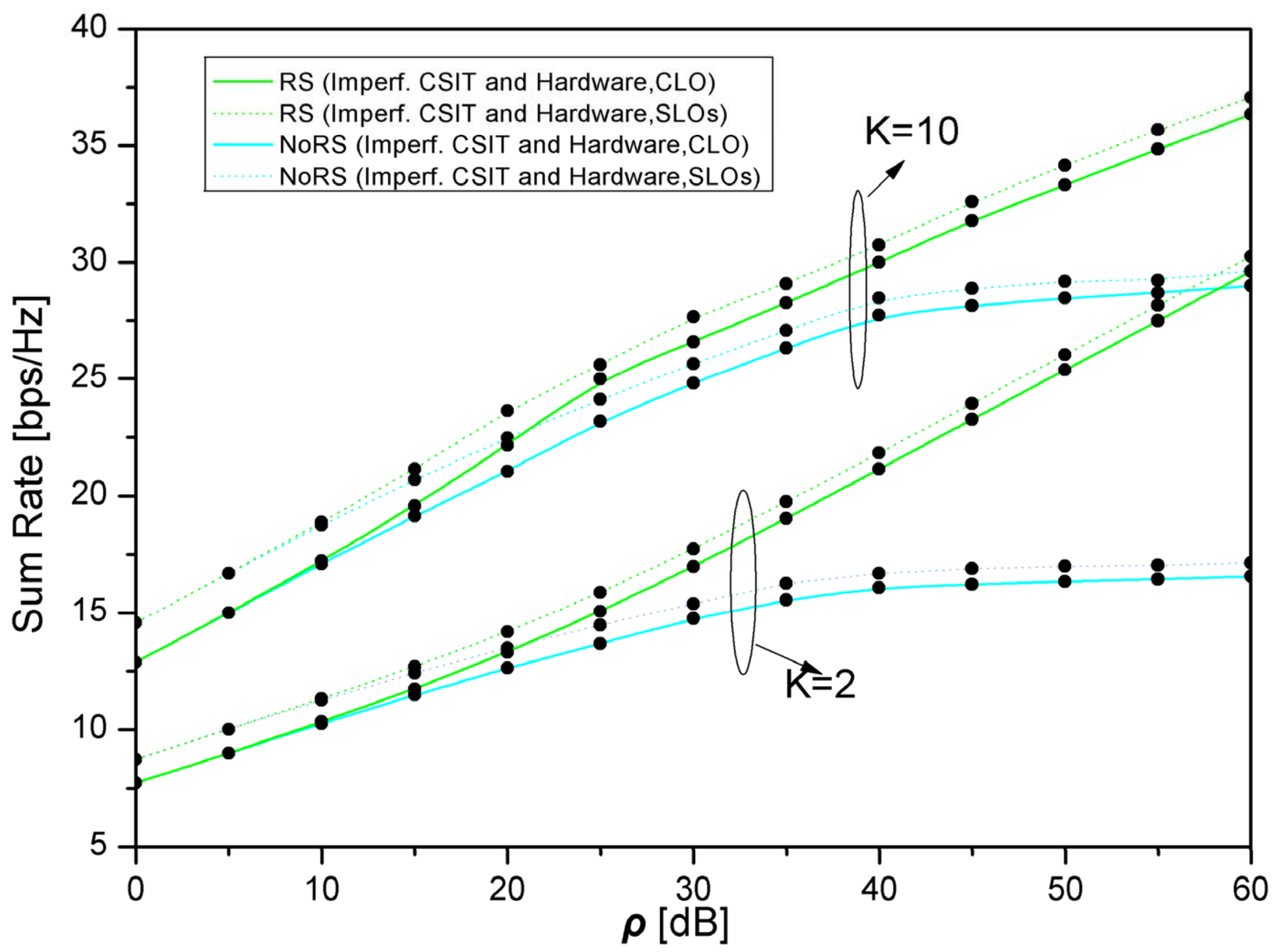}
 \caption{\footnotesize{Sum-rate versus $\rho$ varying $K$ ($M=100$,  $T=500$, $\delta=10^{-4}$, $\kappa^{2}=0$, $\xi_{k}^{\mathrm{UE}}=\sigma^{2}$).}}
 \label{varying}
 \end{center}
 \end{figure}
\subsubsection{Varying the number of users}
The purpose of this subsection is not only to verify that the gain in the achievable rate due to  the common message decreases  with the number
of users~\cite{Dai2016}, but also to quantify the realistic sum-rate under the presence of RTHIs. Given that we have observed the robustness of RS only in the case of multiplicative distortion at the transceiver, in Fig~\ref{varying}, we choose to take into account only the multiplicative impairment of PN. Clearly, increasing the number of UEs from $2$ to $10$, the achievable rate degrades, since the common message has to be now decoded by  more UEs. In such case, Hierarchical-
Rate-Splitting (HRS) has to be introduced~\cite{Dai2016}, in order to retain the benefits of RS. However, HRS accounting for RTHIs is left for future work.
\section{Conclusions}\label{Conclusions}
The rate performance of conventional multi-user broadcasting saturates in both different cases of imperfect CSIT and RTHIs. In fact, a ceiling appears at high SNR. Till now, the RS strategy was proposed to tackle the degradation coming from the multi-user interference induced by imperfect CSIT.   While the RTHIs are inherent in any communication system, their impact on MISO BC channels  was not taken into consideration, since prior work assumed the idealistic scenario of perfect hardware. The objective of this work was to study the  impact of RTHIs in MISO BC channels and investigate the potential robustness of the RS method in both cases of perfect and imperfect CSIT, when the RTHIs are taken into account. Specifically, by generalizing the RS method and existing results to the large system regime, we shed light to the potentials of the RS approach in the cases of additive and multiplicative impairments. 

Initially, we obtained the estimated channel in the realistic scenario where additive and multiplicative impairments are considered. Next, we provided the downlink signal model of a MISO BC channel with  RTHIs. Moreover, we provided the DE analysis of the downlink achievable rate, after designing the precoders for the common and private messages. In particular,  by pursuing a DE analysis we obtained the achievable rates of both common and private parts, in order to investigate the sum-rate achieved by the RS strategy in the practical case with RTHIs. Furthermore, the validation of the analytical results was shown  by means of simulations. Especially, simulations depicted that the asymptotic results can be applicable even for contemporary finite system dimensions of small size. Notably, RS proved to be robust in the case of PN and amplified thermal noise. Unfortunately, when additive impairments are accounted, the rate saturates, if perfect, or more practically, imperfect CSIT is considered. However, it outperforms the NoRS scheme in all cases. Finally, we verified that the gain promised by RS is reduced as the number of UEs increases even in the presence of RTHIs.

\begin{appendices}
\section{Useful Lemmas}
\begin{lemma}[Matrix inversion lemma (I) {\cite[Eq.~2.2]{Bai1}}]\label{lemma:inversion}
\\Let $\bB\in\CC^{M\times M}$ be Hermitian  invertible. Then, for any vector $\bx\in\CC^{M}$, and any scalar $\tau\in\CC^{M}$ such that $\bB+\tau\bx\bx^\H$ is invertible,
\begin{align}
\bx^\H(\bB+\tau\bx\bx^\H)^{-1}=\frac{\bx^\H\bB^{-1}}{1+\tau\bx^\H\bB^{-1}\bx}.\nn
\end{align}
 \end{lemma}

\begin{lemma}[Matrix inversion lemma (II) {\cite[Lemma~2]{Hoydis2013}}]\label{lemma:inversion2}
\\Let $\bB\in\CC^{M\times M}$ be Hermitian  invertible. Then, for any vector $\bx\in\CC^{M}$, and any scalar $\tau\in\CC^{M}$ such that $\bB+\tau\bx\bx^\H$ is invertible,
\begin{align}
(\bB+\tau\bx\bx^\H)^{-1}=\bB-\frac{\bB^{-1}\tau\bx\bx^\H\bB^{-1}}{1+\tau\bx^\H\bB^{-1}\bx}.\nn
\end{align}
 \end{lemma}

\begin{lemma}[Rank-1 perturbation lemma {\cite[Lemma~2.1]{Bai2}}]
\\Let $z\in<0$, $\bB\in\CC^{M\times M}$, $\bB\in\CC^{M\times M}$ with $\bB$ Hermitian  nonnegative-definite, and $\bx\in\CC^{M}$. Then,
\begin{align}
|\tr\left((\bB-z\Id_M)^{-1} -(\bB+\bx\bx^\H-z\Id_M)^{-1}\bB\right)|\leq\frac{\|\bB\|}{|z|}.\nn
\end{align}
 \end{lemma}

\begin{lemma}[{\cite[Lem. B.26]{Bai2010a}}]\label{lemma:asymptoticLimits}
Let $\bB \in \bbC^{M \times M}$ with uniformly bounded spectral norm (with respect to $M$). Consider $\bx$ and $\by$, where $\bx, \by \in \bbC^{M}$, $\bx \sim \cC\cN(\b0, \bPhi_{x})$ and $\by  \sim \cC\cN(\b0, \bPhi_{y})$, are mutually independent and independent of $\bB$. Then, we have
\begin{align}
\frac{1}{M}\bx^{\H}\bB\bx - \frac{1}{M}\tr \bB\bPhi_{x} & \xrightarrow[ M \rightarrow \infty]{\mbox{a.s.}} 0 \label{eq:oneVector}\\
\frac{1}{M}\bx^{\H}\bB\by & \xrightarrow[ M \rightarrow \infty]{\mbox{a.s.}} 0 \label{eq:twoVector}\\
\EE\!\!\left[\left|\left(\frac{1}{M}\bx^{\H}\bB\bx\right)^{2}\!\! - \!\left(\frac{1}{M}\tr \bB \bPhi_{x} \right)^{2} \right|\right] \!\!&\xrightarrow[ M \rightarrow \infty]{\mbox{a.s.}}  0\label{eq:squared}\\
\frac{1}{M^{2}} |\bx^{\H}\bB\by|^{2} - \frac{1}{M^{2}} \tr \bB \bPhi_{x} \bB^{\H} \bPhi_{y}  & \xrightarrow[ M\rightarrow \infty]{\mbox{a.s.}} 0. \label{eq:twoVectorGeneral}
\end{align}
\end{lemma}

\begin{lemma}[{\cite[p. 207]{Tao2012}}]\label{lemma:asymptoticproduct}
Let $\bA\mathrm{,}~\bB\in \mathcal{C}^{M\times M}$ be freely independent random
matrices  with uniformly bounded spectral norm for all $M$. Further, let all the moments of the entries of $\bA\mathrm{,}~\bB$ be finite. Then,
\begin{align}
\frac{1}{M}\tr \bA\bB-\frac{1}{M}\tr \bA \frac{1}{M}\tr \bB \xrightarrow[ M\rightarrow \infty]{\mbox{a.s.}} 0.
\end{align}
\end{lemma}

\begin{Theorem}[{\cite[Theorem 1]{Wagner2012}}]\label{th:detequ}
Let $\bL\in\CC^{M\times M}$ and $\bS\in\CC^{M\times M}$ be Hermitian  nonnegative definite matrices, and let $\bH\in\CC^{M\times K}$ be a random matrix with columns $\bv_k\sim \cC \cN\left(0,\frac{1}{M}\bD_k\right)$. Assume that $\bL$ and the matrices $\bD_k$, $k=1,\dots,K$, have uniformly bounded spectral norms (with respect to $M$). Then, for any $\rho>0$,
\begin{align*}
\frac1M\tr\bL\left(\bH\bH^\H +\bS +\rho\Id_M\right)^{-1} - \frac1M\tr\bL\bT(\rho) \xrightarrow[ M_{\rt} \rightarrow \infty]{\mbox{a.s.}} 0,
\end{align*}
where $\bT(\rho)\in\CC^{M\times M}$ is defined as
$$ \bT(\rho) = \left(\frac1M\sum_{k=1}^K\frac{\bD_k}{1+e_k(\rho)}  +\bS + \rho\Id_M\right)^{-1},$$
and the elements of $\boldsymbol{e}(\rho)=\left[e_1(\rho)\cdots e_K(\rho)\right]^\T$ are defined as $ e_k(\rho)  = \lim_{t \to \infty}e_k^{(t)}(\rho),$ where for $t=1,2,\ldots$
\begin{align}
e_k^{(t)}(\rho)\!=\!\frac1M\tr\bD_k\!\left(\!\!\frac1M\sum_{j=1}^{K}\!\!\frac{\bD_j}{1+e_j^{(t-1)}(\rho)}+\bS+\rho\Id_M\!\!\right)\!\!^{-1}
\end{align}
with initial values $e_k^{(0)}(\rho)=\frac{1}{\rho}$ for all $k$.
\end{Theorem}
\vspace{5pt}

\begin{Theorem}[{\cite[Theorem 2]{Hoydis2013}}]\label{th:detequder}
Let $\bTheta\in\CC^{M\times M}$ be a Hermitian  nonnegative definite matrix with uniformly bounded spectral norm (with respect to $M$). Under the same conditions as in Theorem~\ref{th:detequ}, we have  
\begin{align*}
\frac1M\tr\bL\left(\bH\bH^\H + \bS+\rho\Id_M\right)^{-1}\bK\left(\bH\bH^\H  +\bS+\rho\Id_M\right)^{-1} \nn\\
- \frac1M\tr\bL\bT^{'}(\rho) \xrightarrow[]{\text{a.s.}} 0,
\end{align*}
where $\bT^{'}(\rho)\in\CC^{M\times M}$ is defined as
$$ \bT^{'}(\rho) = \bT(\rho)\bK \bT(\rho) + \bT(\rho)\frac1M\sum_{k=1}^K\frac{\bD_k e^{'}_k(\rho) }{\left(1+e_k(\rho)\right)^2}\bT(\rho)$$
with $\bT(\rho)$ and $\boldsymbol{e}_k(\rho)$ as defined in Theorem~\ref{th:detequ} and $\boldsymbol{e}^{'}(\rho) = \left[e^{'}_1(\rho)\cdots e^{'}_K(\rho)\right]^\T$ given by
\begin{align}
 \boldsymbol{e}^{'}(\rho) &= \left(\Id_K - \bJ(\rho)\right)^{-1}\bv(\rho).
\end{align}
The elements of $\bJ(\rho)\in\CC^{K\times K}$ and $\bv(\rho)\in\CC^{K}$ are defined as
\begin{align}
[\bJ(\rho)]_{kl} = \frac{\frac1M\tr\bD_k\bT(\rho)\bD_l\Tm(\rho)}{M\left(1+e_k(\rho)\right)^2}\nn
\end{align}
and
\begin{align}
[\bv(\rho)]_k = \frac1M\tr\bD_k\Tm(\rho)\bK\Tm(\rho).\nn
\end{align}
\end{Theorem}
\section{Proof of  Theorem~\ref{theorem:RZF}}\label{theorem3}
The proof starts with the derivation of the DE of the normalization parameter $\lambda$. After  making a simple algebraic manipulation to~\eqref{eq:lamda}, we obtain\footnote{It is crucial to mention that the diagonal matrix $\mathrm{diag}\left( \!\hat{\bW}\! \right)$ inside the RZF precoder can be treated as a deterministic matrix with diagonal elements the limits of the individual diagonal elements during the derivation of the DES of the various terms. Specifically, unfolding  the uniform convergence $ \lim \sup_M \max_{1\le i\le M} { |(\bG\bG^{\H})_{mm} - \hat{\bR}_{mm}| } = 0$, we have $\|\frac{1}{M}  \mathrm{diag}(\bG \bG^{H})-\frac{1}{M}\tr\hat{\bR}\| \xrightarrow[ M \rightarrow \infty]{\mbox{a.s.}} 0 $.}
\begin{align}
\lambda=
\frac{K}{\tr  \hat{\bG}^{\H}{ \bSigma}^{2}\hat{\bG} } = \frac{K}{\Psi}.\label{eq:theorem4.6}
\end{align}
We need to define
\begin{align}
 {\bSigma}_{k}&\!\triangleq\!\left(\!\frac{\tilde{\kappa}_{\mathrm{t}_\mathrm{BS}}^{2}}{M}\!\left( \hat{\bW} \!-\!\hat{\bg}_{k}\hat{\bg}_{k}^{\H}\right) \! +\!\frac{\kappa_{\mathrm{r}_\mathrm{UE}}^{2} \mathrm{diag}\left(\! \hat{\bW} \!\right)}{M}\!+\! \frac{\bZ}{M} \!+\!  \al~\! \xi_{k}^{\mathrm{BS}} \Id_M\right)^{\!\!-1}.
\end{align} Next, we can write $\Psi$ in~\eqref{eq:theorem4.6} as
\begin{align}
 \Psi&=\sum_{k=1}^{K}\hat{\bg}_{k}^{\H} \bSigma^{2}\hat{\bg}_{k}\\
 &\asymp\frac{1}{M}\sum_{k=1}^{K}\frac{\frac{1}{M}\tr \hat{\bR}_{k}\bSigma^{2}_{k}}{\left( 1+\frac{1}{M}\tr \hat{\bR}_{k}\bSigma_{k} \right)^{2}}\label{lambda1}\\
 &\asymp\frac{1}{M}\sum_{k=1}^{K}\frac{{\delta}_{k}^{'}}{\left( 1+{\delta}_{k} \right)^{2}},\label{lambda2}
\end{align}
where we have applied Theorems~\ref{th:detequ} and~\ref{th:detequder} for $\bL=\hat{\bR}_{k}$ and $\bK=\Id_M$. Also, we have denoted ${\delta}_{k}=\frac{1}{M}\tr \hat{\bR}_{k}\bT$ and $\delta_{k}^{'}=\frac{1}{M}\tr \hat{\bR}_{k}\hat{\bT}^{'}$. Hence, ${\lambda}\asymp \bar{\lambda} $.
The rest part of the desired signal power can be expressed as
\begin{align}
\bg_{k,n}^{\H}\bff_{k}
&=\tilde{\bg}_{k}^{\H}\widetilde\bTheta_{k,n}\bSigma \hat{\bg}_{k}.\label{desired1} 
\end{align}
In~\eqref{desired1}, we have substituted the RZF precoder. Dividing by $\frac{1}{M}$, we obtain
\begin{align}
\frac{1}{M}\tilde{\bg}_{k}^{\H}\widetilde\bTheta_{k,n}\bSigma \hat{\bg}_{k}
&=\frac{1}{M} \hat{\bg}_{k}^{\H}\widetilde\bTheta_{k,n}\bSigma \hat{\bg}_{k}\label{desired2} \\
&=\frac{\frac{1}{M}\tr \widetilde\bTheta_{k,n}\bSigma_{k} \hat{\bR}_{k}}{1+\frac{1}{M}\tr\bSigma_{k} \hat{\bR}_{k}}\label{desired3}\\
&=\frac{\frac{1}{M}\tr \widetilde\bTheta_{k,n}\frac{1}{M}\tr\bSigma_{k} \hat{\bR}_{k}}{1+\frac{1}{M}\tr\bSigma_{k} \hat{\bR}_{k}}\label{desired4}\\
&=\frac{\frac{1}{M}\tr \widetilde\bTheta_{k,n}\frac{1}{M}\tr\hat{\bR}_{k} \bT }{1+\frac{1}{M}\tr\hat{\bR}_{k}\bT }.\label{desired5}
\end{align}
Herein, we have applied  Lemmas~\ref{lemma:inversion},~\ref{lemma:asymptoticLimits}, and \ref{lemma:asymptoticproduct} in \eqref{desired2} and \eqref{desired3}, respectively. Moreover, we have exploited Theorem~\ref{th:detequ} in  \eqref{desired4} for $\bL=\hat{\bR}_{k}$. 
Writing more concisely the last equation, we obtain
\begin{align}
\frac{1}{M}\bg_{k}^{\H}\widetilde\bTheta_{k,n}\bSigma \hat{\bg}_{k}&=\frac{\frac{1}{M}\tr \widetilde\bTheta_{k,n}{\delta}_{k}}{1+{\delta}_{k} }.\label{desired6}
\end{align}
Note that the manipulation regarding $\widetilde\bTheta_{k,n}$ follows a similar analysis to~\cite{Papazafeiropoulos2016}.
The proof continues with the derivation of each term of the interference part of $\mathrm{SINR}_{k}^{\mathrm{p}}$
\begin{align}
\frac{\rho_{j}}{K}\sum_{j\ne k}^{K}|\tilde{\bg}_{k}^{\H}\widetilde\bTheta_{k,n}\bff_{j}|^2&+\EE\left[ |\tilde{\bg}_{k}^{\H}\widetilde\bTheta_{k,n}\etv_{\mathrm{t},n}^{\mathrm{BS}} |^2\right] \nn\\
&+\EE\left[ |\eta_{\mathrm{r},n}^{\mathrm{UE}}|^2\right] +\xi_{k}^{\mathrm{UE}}.
\end{align}
Hence, making use of~\eqref{current} to the first term, we obtain by means of Lemmas~\ref{lemma:asymptoticLimits} and~\ref{lemma:inversion}
\begin{align}
\frac{1}{M^{2}}|\tilde{\bg}_{k}^{\H}\widetilde\bTheta_{k,n}\bff_{j}|^2&=\frac{1}{M^{2}}|\hat{\bg}_{k}^{\H}\widetilde\bTheta_{k,n}\bSigma \hat{\bg}_{j}|^2\label{Int1} \\
&=\frac{1}{M^{2}}\frac{\hat{\bg}_{k}^{\H}\widetilde\bTheta_{k,n}\bSigma_{j} \hat{\bg}_{j}\hat{\bg}^{\H}_{j}\bSigma_{j}\widetilde\bTheta_{k,n}^{*}\hat{\bg}_{k}}{\left( 1+\hat{\bg}_{j}^{\H}\bSigma_{j} \hat{\bg}_{j} \right)^2}\label{Int2}\\
&=\frac{1}{M^{2}}\frac{\hat{\bg}_{k}^{\H}\widetilde\bTheta_{k,n}\bSigma_{j} \hat{\bR}_{j}\bSigma_{j}\widetilde\bTheta_{k,n}^{*}\hat{\bg}_{k}}{\left( 1+\hat{\bg}_{j}^{\H}\bSigma_{j} \hat{\bg}_{j} \right)^2}\label{Int3},
\end{align}
where in~\eqref{Int3}, we have taken into consideration that $\hat{\bg}_{k} $ and $\hatvg_{j} $ are mutually independent. Given that $\bSigma_{j}$  is not independent of $\hat{\bg}_{k}$, we employ  Lemma~\ref{lemma:inversion2}, which yields
\begin{align}
\bSigma_j={\bSigma}_{jk}-\frac{{\bSigma}_{jk}\hatvg_{k}\hatvg_{k}^{\H}{\bSigma}_{jk}}{1+\hatvg^\H_{k}  {\bSigma}_{jk}\hatvg_{k}}\label{eq:theorem2.I.51}, 
\end{align}
where the new matrix ${\bSigma}_{jk}$ is defined as
\begin{align}
{\bSigma}_{jk}\!=\!
\left(\frac{\tilde{\kappa}_{\mathrm{r}_\mathrm{BS}}^{2}}{M}\left( \hat{\bW} \!-\!\hat{\bg}_{k}\hat{\bg}_{k}^{\H}\!-\!\hat{\bg}_{j}\hat{\bg}_{j}^{\H}\right)  +\!\frac{\kappa_{\mathrm{r}_\mathrm{UE}}^{2} \mathrm{diag}\left(\! \hat{\bW} \!\right)}{M}\right.\nn\\
\left.+\!  \frac{\bZ}{M} \!+\!  \al~\! \xi_{k}^{\mathrm{UE}} \Id_M\right)^{-1}.
\end{align}
After substituting~\eqref{eq:theorem2.I.51} into~\eqref{Int3}, we have
\begin{align}
  \frac{1}{M^{2}}\left|\hat{\bg}^\H_{k,n}{\bSigma} \hatvg_{j,n} \right|^{2} 
&=\frac{{Q}_{jk}}{M\left(1+{\delta_{j}}\right)^{2}},\label{eq:theorem2.I.6}
\end{align}           
where ${Q}_{jk}$ is given by~\eqref{eq:theorem2.I.mu1}.
\begin{figure*}
\begin{align}
{Q}_{jk}&= \hat{\bg}^\H_{k}\widetilde\bTheta_{k,n} \bSigma_{jk}\hat{\bR}_{j} \bSigma_{jk} \widetilde\bTheta_{k,n}\hat{\bg}_{k}\!+\!\frac{\left|  \hat{\bg}^\H_{k}\widetilde\bTheta_{k,n}\bSigma_{jk} \hat{\bg}_{k}\right|^{2}\hat{\bg}^\H_{k} \bSigma_{jk}\hat{\bR}_{j}\bSigma_{jk}\hat{\bg}_{k}}{\left( 1+\hatvg^\H_{k} \bSigma_{jk} \hatvg_{k} \right)^{2}}\nn\\
&-2\mathrm{Re}\left\{  \frac{\hatvg^\H_{k}\bSigma_{jk}\widetilde\bTheta_{k,n} \bg_{k}\bg_{k}^{\H}\bSigma_{jk}\hat{\bR}_{j}\bSigma_{jk}\hatvg_{k}}{1+\hatvg^\H_{k} \bSigma_{jk} \hatvg_{k}}\right\}.
 \label{eq:theorem2.I.mu1}
\end{align}
\line(1,0){470}
\end{figure*}
We proceed with the derivation of the DE of each term in~\eqref{eq:theorem2.I.mu1}. Specifically, we have
\begin{align}
 \frac{1}{M^{2}}\hat{\bg}^\H_{k}\widetilde\bTheta_{k,n} \bSigma_{jk}\hat{\bR}_{j} \bSigma_{jk} \widetilde\bTheta_{k,n}^{*}\hat{\bg}_{k}
 &\asymp  \frac{1}{M^{2}}\tr \hat{\bR}_{k}\bSigma_{jk}\hat{\bR}_{j}\bSigma_{jk}\\
 &\asymp\frac{1}{M^{2}}\tr \hat{\bR}_{j}\hat{\bT}^{''}\\
 &=\frac{ \delta_{j}^{''}}{M},
\end{align}
where we have applied Lemma~\ref{lemma:asymptoticLimits} and Theorem~\ref{th:detequder} for $\bL=\hat{\bR}_{j}$ and $\bK=\hat{\bR}_{k}$.
Similarly, we have
\begin{align}
\frac{1}{M^{2}}\left|  \hat{\bg}^\H_{k}\widetilde\bTheta_{k,n}\bSigma_{jk} \hat{\bg}_{k}\right|^{2}
&\asymp \frac{1}{M^{2}}\tr\widetilde\bTheta_{k,n}\hat{\bR}_{k}\bSigma_{jk}\hat{\bR}_{k}\widetilde\bTheta_{k,n}^{*}\bSigma_{jk}\nn\\
&= \frac{1}{M^{2}}\tr\hat{\bR}_{k}\bSigma_{jk}\hat{\bR}_{k}\bSigma_{jk}\\
&\asymp \frac{1}{M^{2}}\tr\hat{\bR}_{k}\hat{\bT}^{''}\\
&=\frac{ \delta_{k}^{''}}{M},
\end{align}
where $\bL=\hat{\bR}_{k}$ and $\bK=\hat{\bR}_{k}$, and  $\delta_{k}^{''}=\frac{1}{M}\tr \hat{\bR}_{k}\hat{\bT}^{''}$.
Furthermore, we have
\begin{align}
\frac{1}{M^{2}}\hat{\bg}^\H_{k} \bSigma_{jk}\hat{\bR}_{j}\bSigma_{jk}\hat{\bg}_{k}&\asymp \frac{1}{M^{2}}\tr\hat{\bR}_{k}\bSigma_{jk}\hat{\bR}_{j}\bSigma_{jk}\\
&=\frac{ \delta_{k}^{'}}{M},
\end{align}
where we make use of Theorems~\ref{th:detequ} and~\ref{th:detequder} as well as Lemmas~1 and~\ref{lemma:asymptoticLimits} as before. The next term is written as
\begin{align}
\frac{1}{M^{2}}\hatvg^\H_{k}\bSigma_{jk}\widetilde\bTheta_{k,n} \bg_{k} &\asymp \frac{1}{M}\tr\widetilde\bTheta_{k,n} \frac{1}{M}\tr\hat{\bR}_{k}\bSigma_{jk}\label{int4} \\
&\asymp \frac{1}{M}\tr\widetilde\bTheta_{k,n} \frac{1}{M}\tr\hat{\bR}_{k}\bT\label{int5}\\
&= \frac{1}{M}\tr\widetilde\bTheta_{k,n} {\delta}_{k}\label{int6}, 
\end{align}
where in~\eqref{int4}, we have applied both Lemmas~\ref{lemma:asymptoticLimits} and~\ref{lemma:asymptoticproduct}, while in the next equation, we have applied Theorem~\ref{th:detequ}. Hence,~\eqref{eq:theorem2.I.mu1} becomes
\begin{align}
\!\!{Q}_{jk}\!\asymp\! \frac{ \delta_{j}^{''}}{M}\!\!+\!\frac{\left|{\delta_{k}^{''}}\right|^{2}\delta_{k}^{''}}{M\left( 1\!+\!\delta_{j} \right)^{2}}\!-\!2\mathrm{Re}\left\{ \! \frac{\frac{1}{M}\tr\widetilde\bTheta_{k,n} {\delta}_{k}\delta_{k}^{''} }{M\left( 1\!+\!\delta_{j} \right)}\!\right\}\!,
 \label{eq:theorem2.I.mu}
\end{align}
Moreover, the other terms corresponding to the transmit and receive distortions become
\begin{align}
\frac{1}{M}\EE\left[ |\tilde{\bg}_{k}^{\H}\widetilde\bTheta_{k,n}\etv_{\mathrm{t},n}^{\mathrm{BS}} |^2\right]&=\frac{1}{M}\bg_{k}^{\H}\widetilde\bTheta_{k,n} \bm \Lambda^{\mathrm{BS}}_{n} \widetilde\bTheta_{k,n} \bg_{k}\nn\\
&\asymp \rho
\kappa_{\mathrm{t}_\mathrm{BS}}^{2}  \frac{1}{M}\tr{\bR}_{k}\\
\EE\left[ |\eta_{\mathrm{r},n}^{\mathrm{UE}}|^2\right] &=\frac{1}{M}\Upsilon^{\mathrm{UE}}_{n}\nn\\
&\asymp \rho \kappa_{\mathrm{r}_\mathrm{UE}}^{2}\frac{1}{M}\tr{\bR}_{k}.
\end{align}
Therefore,  the proof for the derivation of   $\mathrm{SINR}_{k}^{\mathrm{p}}$ is concluded.
As far as the desired signal part of the SINR describing the common message is concerned, we have
\begin{align}
 \frac{1}{M}\bg_{k,n}^{\H} \bff_{c}&=\alpha_{k} \frac{1}{M}\tilde{\bg}_{k}^{\H}\widetilde\bTheta_{k,n}\hat{\bg}_{k}\\
  &=\alpha_{k} \frac{1}{M}\tr\widetilde\bTheta_{k,n} \frac{1}{M}\tr \hat{\bR}_{k},
\end{align}
where  we first used Lemma~\ref{lemma:asymptoticLimits}, and then Lemma~\ref{lemma:asymptoticproduct}.
\section{Proof of Proposition~\ref{prop:inequality}}\label{proofinequality}
Let us denote $\bar{Y}=\frac{\rho t}{K M}\left({\bar{\lambda}}\!\sum_{j\ne k}^{K}\!\frac{{Q}_{jk}}{\left(1+{\delta_{j}}\right)^{2}\!}+ \bar{m}_{k}\right.$ $\left.\left(   \kappa_{\mathrm{t}_\mathrm{BS}}^{2}+ \kappa_{\mathrm{r}_\mathrm{UE}}^{2} \right)\right) +\xi_{k}^{\mathrm{UE}}$ with $\bar{m}_{k}=\tr \hat{\bR}_{k}$. In the case $\bar{Y}>1$, the private part of RS achieves the same sum-rate as the conventional multi-user BC with full power. In other words, the equality in~\eqref{inequality} nearly holds. Having in mind that the common message should be decoded by all UEs, it is reasonable to   allocate less power to the common message,  as their number increases because the rate of the common decreases, i.e., the benefit of common message reduces. Thus, during the power allocation for the common message, the number of UEs should be taken into account. Moreover, following a similar rationale as in~\cite{Dai2016}, we set $\bar{Y}>K$.  We   have 
\begin{align}
\!\!\!t=\frac{K^{2}M }{{\bar{\lambda}}\!\sum_{j\ne k}^{K}\!\frac{\rho {Q}_{jk}}{\left(1+{\delta_{j}}\right)^{2}\!}\!+\!\bar{m}_{k}\rho \left(   \kappa_{\mathrm{t}_\mathrm{BS}}^{2}\!+\! \kappa_{\mathrm{t}_\mathrm{UE}}^{2} \right)+K M \xi_{k}^{\mathrm{UE}}}.\label{optimal} 
\end{align}
By choosing  $t$ as the smaller value between~\eqref{optimal} and   $1$, the inequality in~\eqref{inequality} becomes equality.
In the low-SNR regime $\rho \to 0$, \eqref{tau} gives $t=1$. In other words,
transmission of the common message is not beneficial at this regime. However, increasing the SNR, the transmission of the common message enhances the sum-rate, when the sum-rate due to only private messages tends to saturate. Upper bounding the rate loss between the private messages of the NoRS and RS, we obtain similar to~\cite{Dai2016}
\begin{align}
\sum_{j=1}^{K} \left( \mathrm{R}_{j}^{\mathrm{NoRS}}-\mathrm{R}_{j}^{\mathrm{p}} \right)
&\le \log_2 e.
\end{align}

\end{appendices}                                                                                                                                                                                                                                                                        
\bibliographystyle{IEEEtran}

\bibliography{mybib}
\begin{IEEEbiography}[{\includegraphics[width=1in,height=1.25in,clip,keepaspectratio]{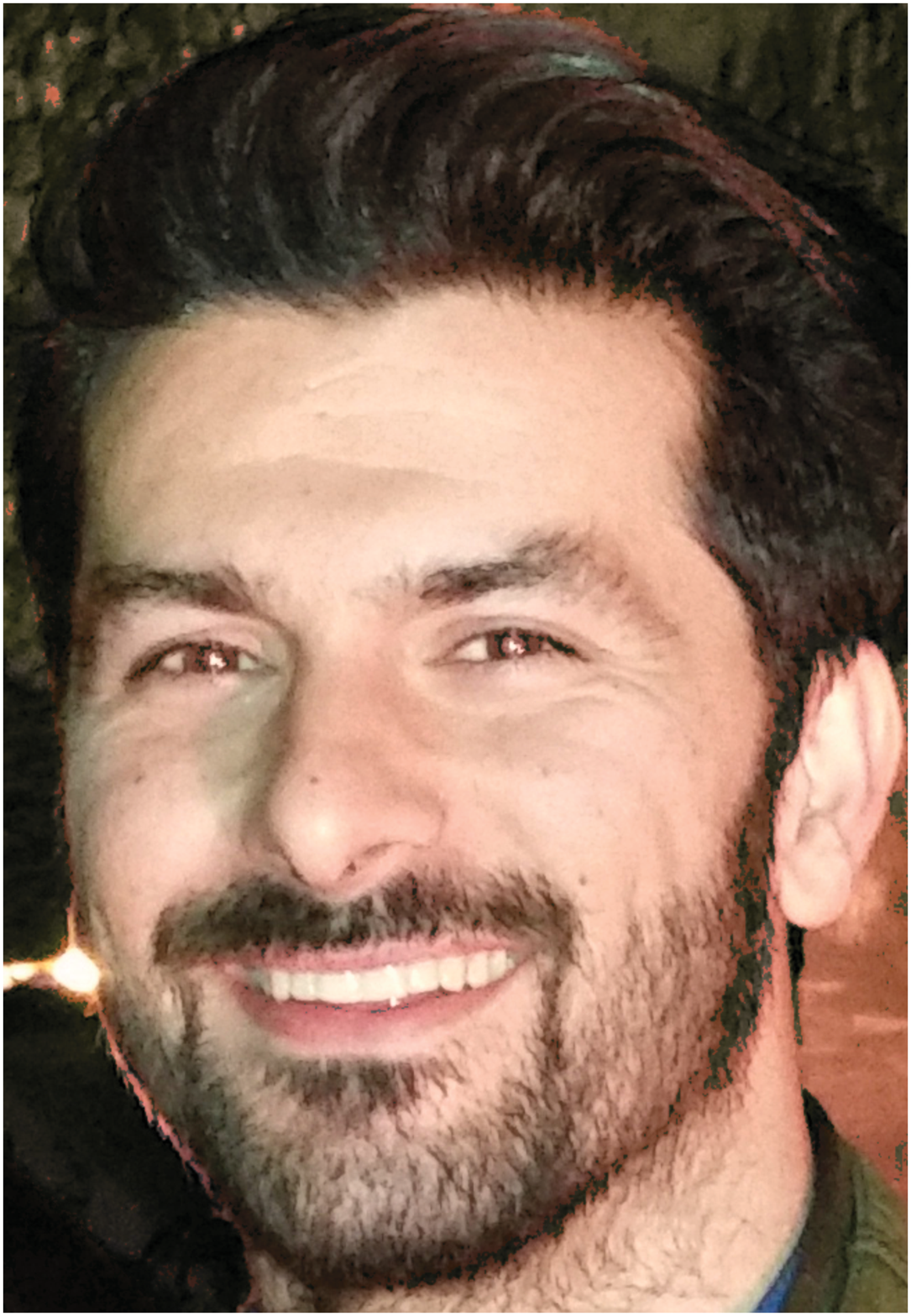}}]{Anastasios Papazafeiropoulos}[S'06-M'10] is currently a Research Fellow in IDCOM at the University of Edinburgh, U.K. He obtained the B.Sc in Physics and the M.Sc. in Electronics and Computers science both with distinction from the University of Patras, Greece in 2003 and 2005, respectively. He then received the Ph.D. degree from the same university in 2010. From November 2011 through December 2012 he was with the Institute for Digital Communications (IDCOM) at the University of Edinburgh, U.K. working as a postdoctoral Research Fellow, while during 2012-2014 he was a Marie Curie Fellow at Imperial College London, U.K.   Dr. Papazafeiropoulos has been involved in several EPSCRC and EU FP7 HIATUS and HARP projects. His research interests span massive MIMO, 5G wireless networks, full-duplex radio, mmWave communications, random matrices theory, signal processing for wireless communications, hardware-constrained communications,
and performance analysis of fading channels. 
\end{IEEEbiography}
\begin{IEEEbiography}
[{\includegraphics[width=1in,height=1.25in,clip,keepaspectratio]{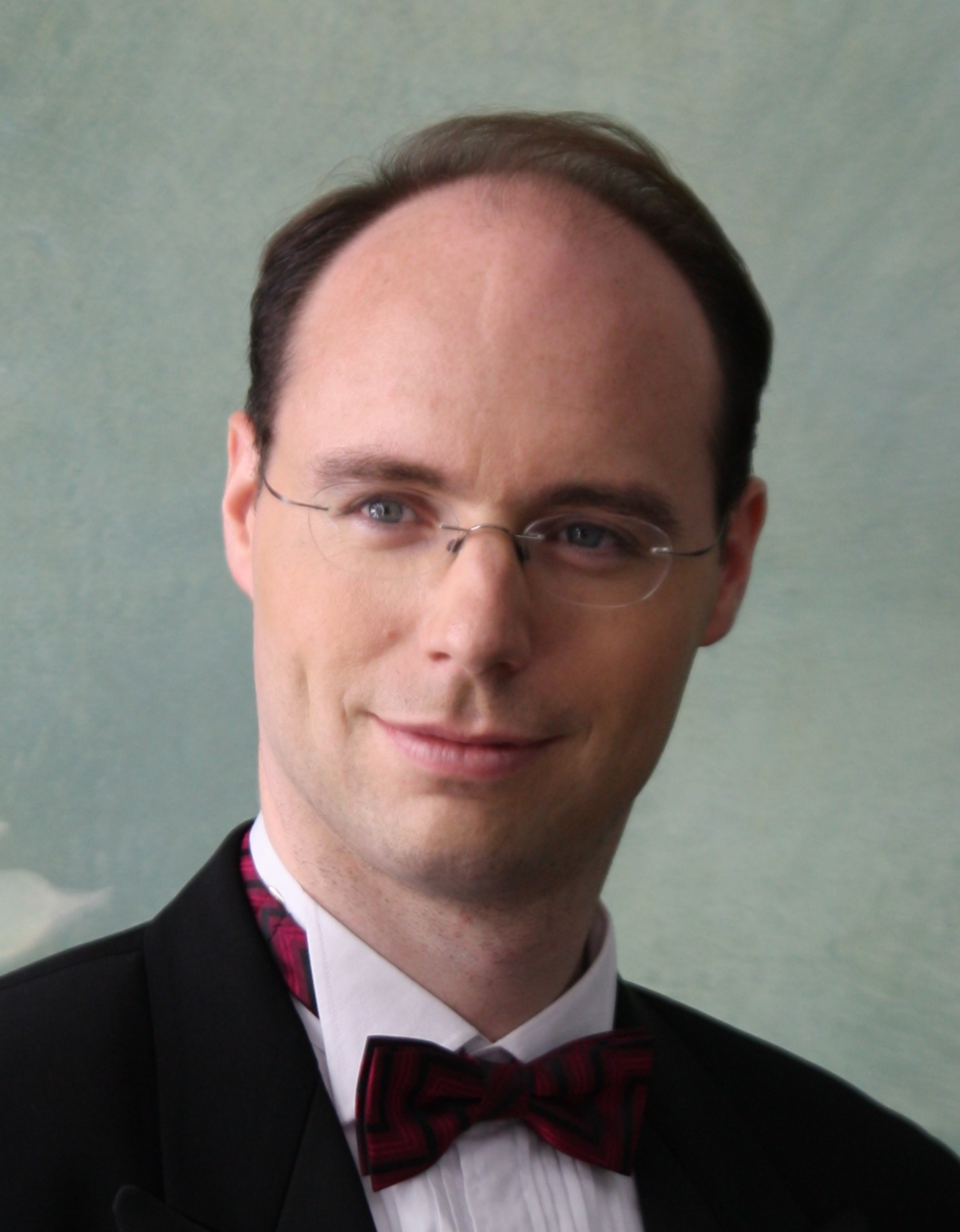}}]{Bruno Clerckx}[]  is a Senior Lecturer (Associate Professor) in the Electrical and Electronic Engineering Department at Imperial College London (London, United Kingdom). He received his M.S. and Ph.D. degree in applied science from the Université catholique de Louvain (Louvain-la-Neuve, Belgium) in 2000 and 2005, respectively. From 2006 to 2011, he was with Samsung Electronics (Suwon, South Korea) where he actively contributed to 3GPP LTE/LTE-A and IEEE 802.16m and acted as the rapporteur for the 3GPP Coordinated Multi-Point (CoMP) Study Item. Since 2011, he has been with Imperial College London, first a Lecturer and now as a Senior Lecturer. From March 2014 to March 2016, he also occupied an Associate Professor position at Korea University, Seoul, Korea. He also held visiting research appointments at Stanford University, EURECOM, National University of Singapore and The University of Hong Kong.

He is the author of 2 books, 120 peer-reviewed international research papers, 150 standard contributions and the inventor of 75 issued or pending patents among which 15 have been adopted in the specifications of 4G (3GPP LTE/LTE-A and IEEE 802.16m) standards. Dr. Clerckx served as an editor for IEEE TRANSACTIONS ON COMMUNICATIONS from 2011-2015 and is currently an editor for IEEE TRANSACTIONS ON WIRELESS COMMUNICATIONS. He is an Elected Member of the IEEE Signal Processing Society SPCOM Technical Committee. His research area is communication theory and signal processing for wireless networks.
\end{IEEEbiography}
\begin{IEEEbiography}
[{\includegraphics[width=1in,height=1.25in,clip,keepaspectratio]{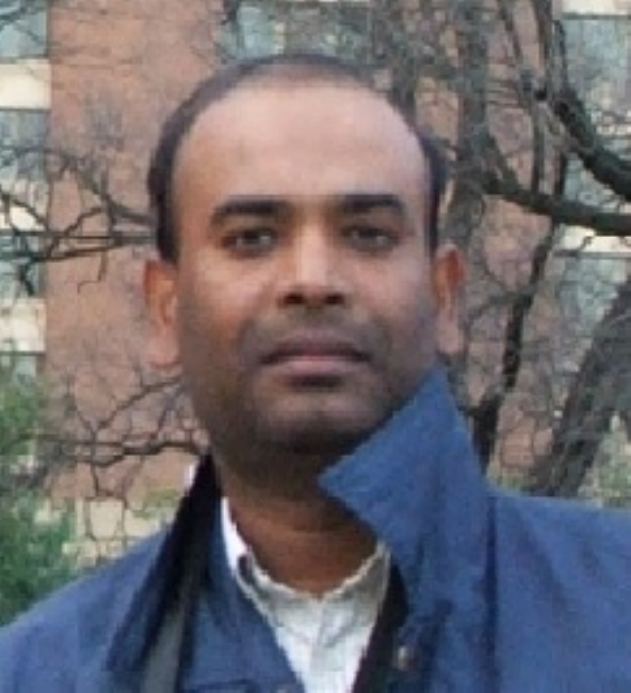}}]{Tharmalingam Ratnarajah}[A'96-M'05-SM'05] is currently with the Institute for Digital Communications, University of Edinburgh, Edinburgh, UK, as a Professor in Digital Communications and Signal Processing and the Head of Institute for Digital Communications. His research interests include signal processing and information theoretic aspects of 5G and beyond wireless networks, full-duplex radio, mmWave communications, random matrices theory, interference alignment, statistical and array signal processing and quantum information theory. He has published over 300 publications in these areas and holds four U.S. patents. 
He was the coordinator of the FP7 projects ADEL (3.7M\euro) in the area of licensed shared access for 5G wireless networks and HARP (3.2M\euro) in the area of highly distributed MIMO and FP7 Future and Emerging Technologies projects HIATUS (2.7M\euro) in the area of interference alignment and CROWN (2.3M\euro) in the area of cognitive radio networks. Dr Ratnarajah is a Fellow of Higher Education Academy (FHEA), U.K..
\end{IEEEbiography}

\end{document}